\documentclass[prd,preprint,tightenlines,showpacs,preprintnumbers,nofootinbib,eqsecnum,superscriptaddress]{revtex4}

\usepackage{dcolumn}
\usepackage{bm}
\usepackage{epsfig}
\usepackage{amsfonts}
\usepackage{amssymb,amscd}
\usepackage{tabularx}
\usepackage{booktabs,multirow}

\def\lsim{\raise0.3ex\hbox{$<$\kern-0.75em\raise-1.1ex\hbox{$\sim$}}}

\def\gsim{\raise0.3ex\hbox{$>$\kern-0.75em\raise-1.1ex\hbox{$\sim$}}}

\newcommand{\be}{\begin{equation}}

\newcommand{\ee}{\end{equation}}

\def\beq{\begin{equation}}

\def\eeq{\end{equation}}

\def\beqa{\begin{eqnarray}}

\def\eeqa{\end{eqnarray}}

\newcommand{\ba}{\begin{eqnarray}}

\newcommand{\ea}{\end{eqnarray}}

\def\gappeq{\mathrel{\rlap {\raise.5ex\hbox{$>$}}

{\lower.5ex\hbox{$\sim$}}}}

\def\lappeq{\mathrel{\rlap{\raise.5ex\hbox{$<$}}

{\lower.5ex\hbox{$\sim$}}}}

\def\Toprel#1\over#2{\mathrel{\mathop{#2}\limits^{#1}}}

\begin{document}

\title{Investigating nuclear effects in lepton - ion DIS at the LHC}

\author{Reinaldo {\sc Francener}}
\email{reinaldofrancener@gmail.com}
\affiliation{Instituto de Física Gleb Wataghin - Universidade Estadual de Campinas (UNICAMP), \\ 13083-859, Campinas, SP, Brazil. }

\author{Victor P. {\sc Gon\c{c}alves}}
\email{barros@ufpel.edu.br}
\affiliation{Institute of Physics and Mathematics, Federal University of Pelotas, \\
  Postal Code 354,  96010-900, Pelotas, RS, Brazil}

\author{Diego R. {\sc Gratieri}}
\email{drgratieri@id.uff.br}
\affiliation{Escola de Engenharia Industrial Metal\'urgica de Volta Redonda,
Universidade Federal Fluminense (UFF),\\
 CEP 27255-125, Volta Redonda, RJ, Brazil}
\affiliation{Instituto de Física Gleb Wataghin - Universidade Estadual de Campinas (UNICAMP), \\ 13083-859, Campinas, SP, Brazil. }

\begin{abstract}
Recent studies have demonstrated that the far-forward physics program of the Large Hadron Collider (LHC) can be useful to probe the hadron structure with GeV-TeV neutrinos and muons. In particular, these studies indicate that the measurement of the   muon - ion and neutrino - ion cross - sections by the same experiment is feasible.  In this paper, we investigate the impact of nuclear effects on the muon - tungsten ($\mu W$) and neutrino - tungsten  ($\nu W$) deep inelastic scattering (DIS) events at FASER$\nu$ and its proposed upgrade FASER$\nu 2$. We estimate the rates associated with the inclusive cross - sections and for events with a charm tagged in the final state  considering different parameterizations for the nuclear parton distribution functions.  These results point out that muon and neutrino - induced interactions probe complementary kinematical ranges and that a simultaneous analysis of associated events will allow to test the universality (or not) of the nuclear effects. Moreover,  we propose the study of  the ratio between the charm tagged and inclusive events in order to discriminate between the distinct modeling of the nuclear effects at small - $x$. Our results indicate that a future experimental reconstruction of $\mu W$ and $\nu W$ DIS events at the LHC is  a promising  way to improve our understanding of nuclear effects and decrease the current uncertainties in parton distribution functions.
\end{abstract}



\maketitle

\vspace{1cm}

\section{Introduction}

\begin{figure}[t]
	\centering
	\begin{tabular}{ccc}
	\includegraphics[width=0.33\textwidth]{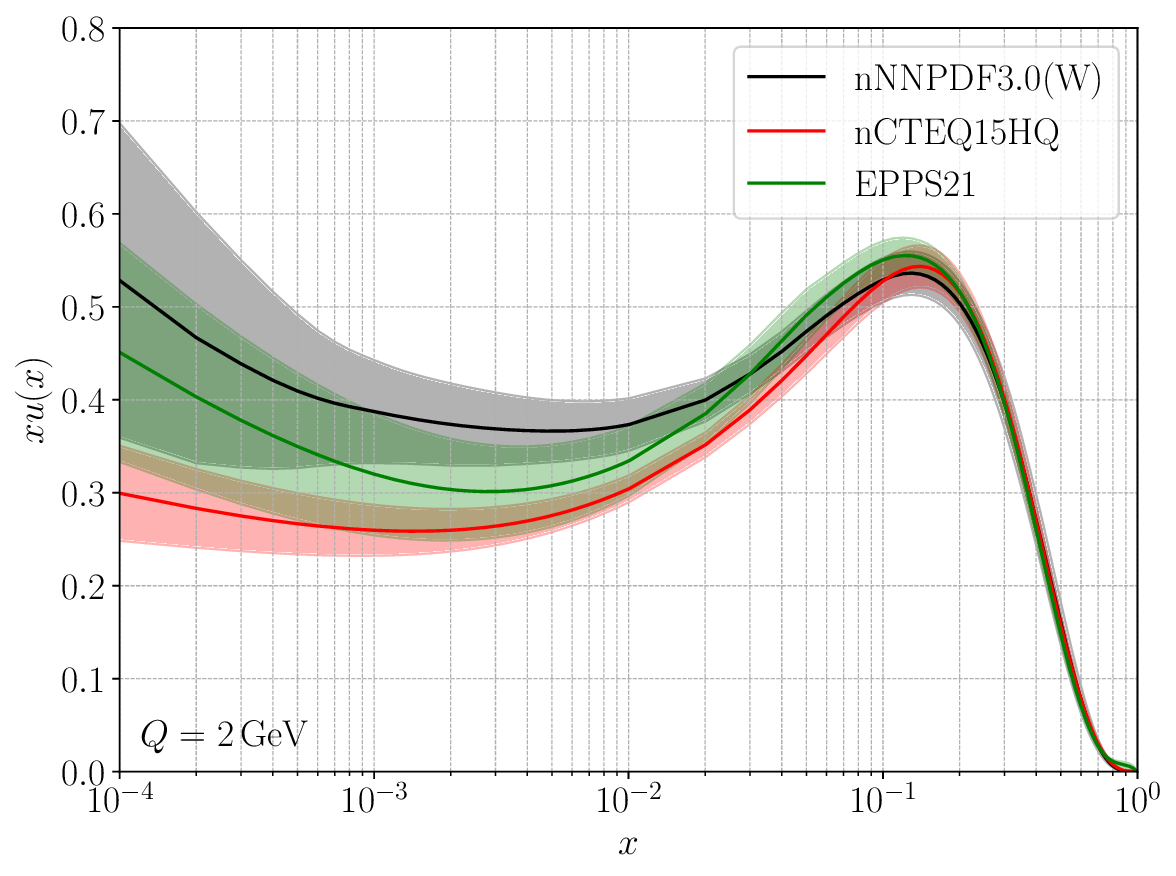}
	\includegraphics[width=0.33\textwidth]{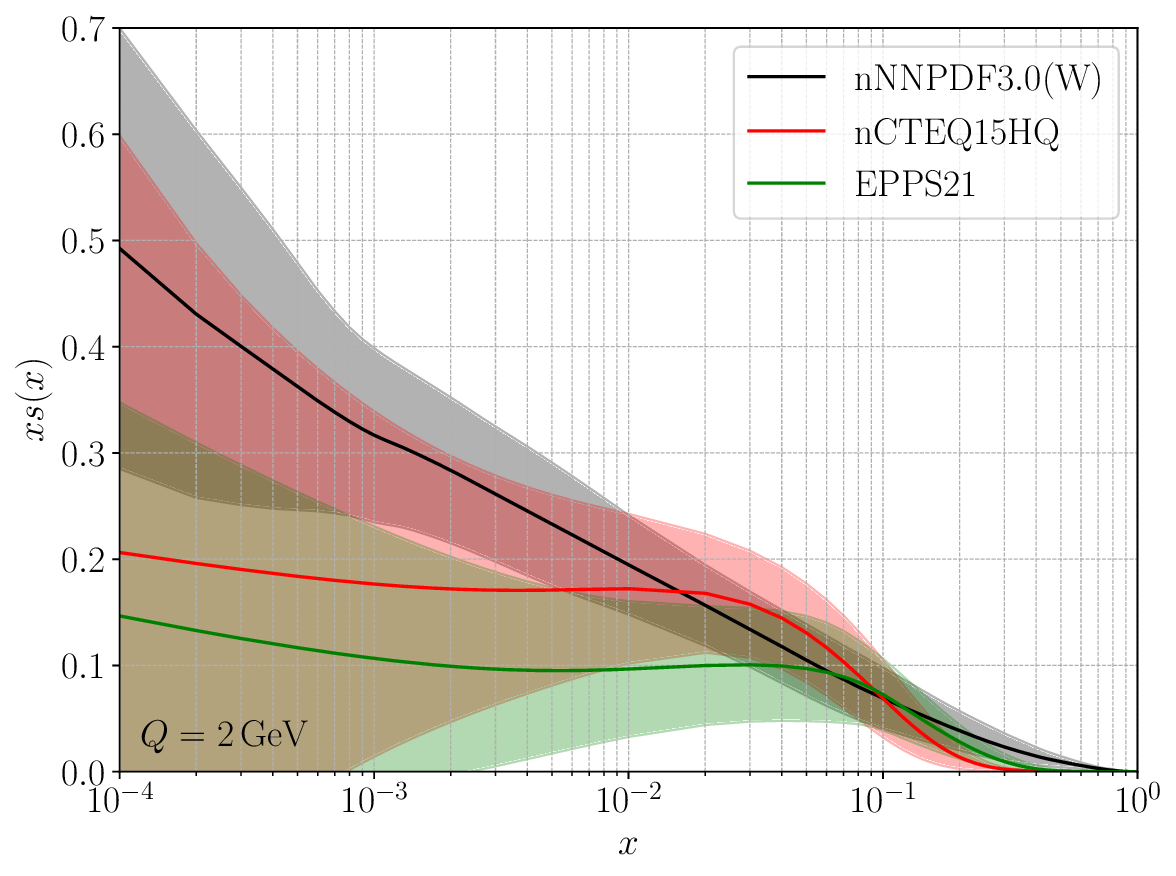}
	\includegraphics[width=0.33\textwidth]{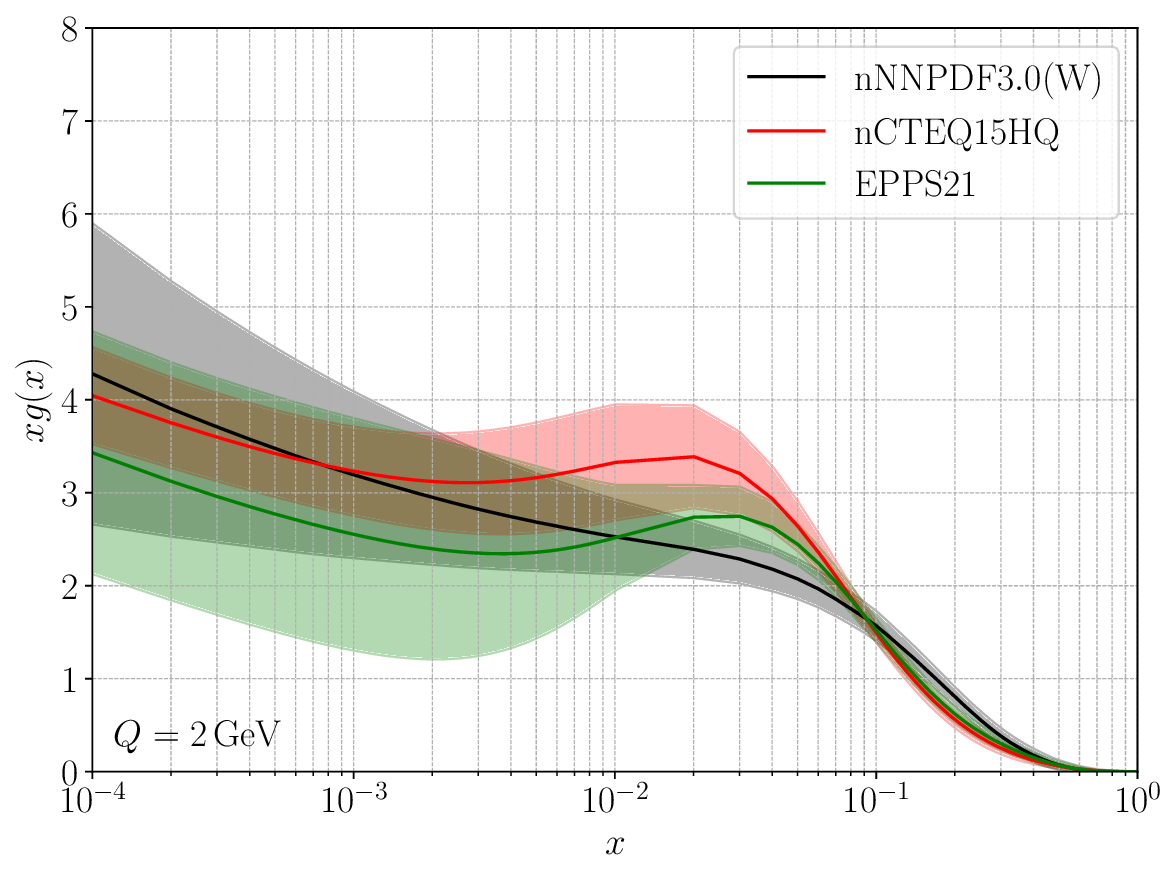}  \\
    \includegraphics[width=0.33\textwidth]{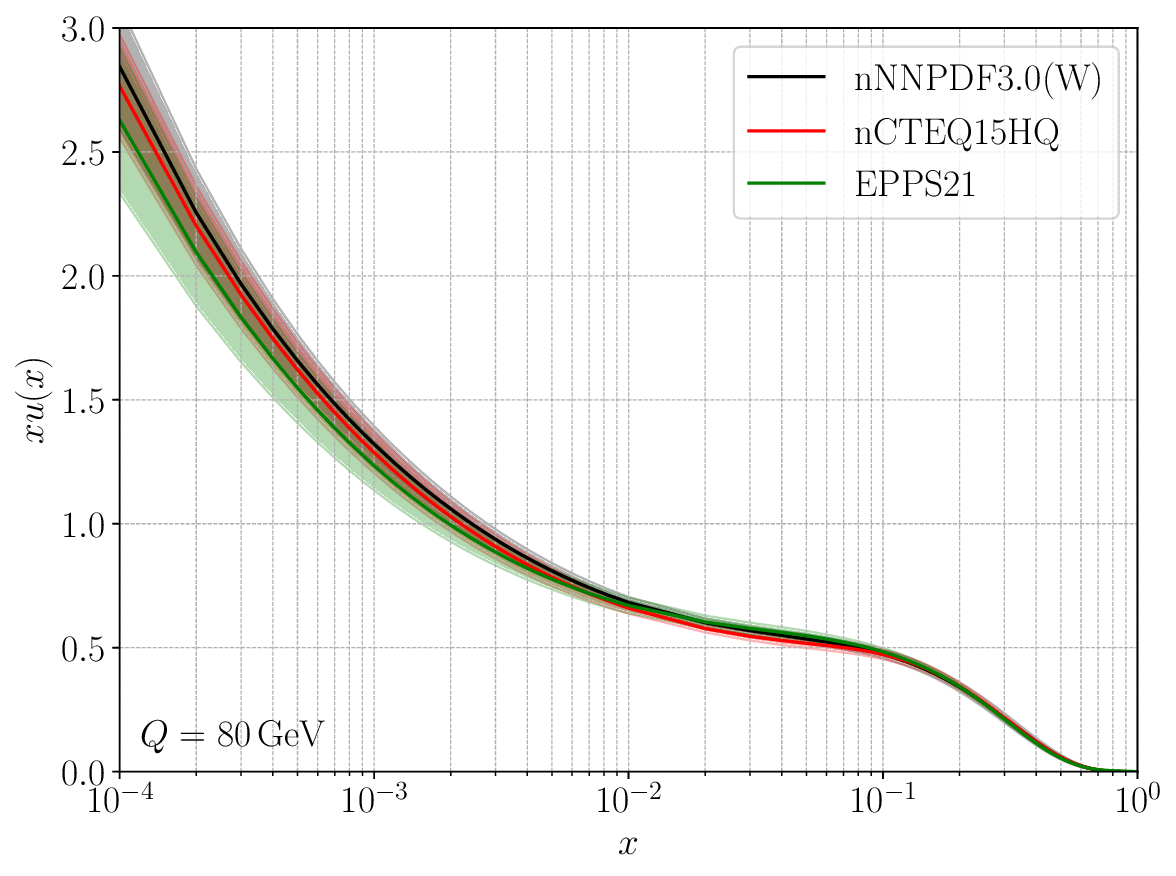}
	\includegraphics[width=0.33\textwidth]{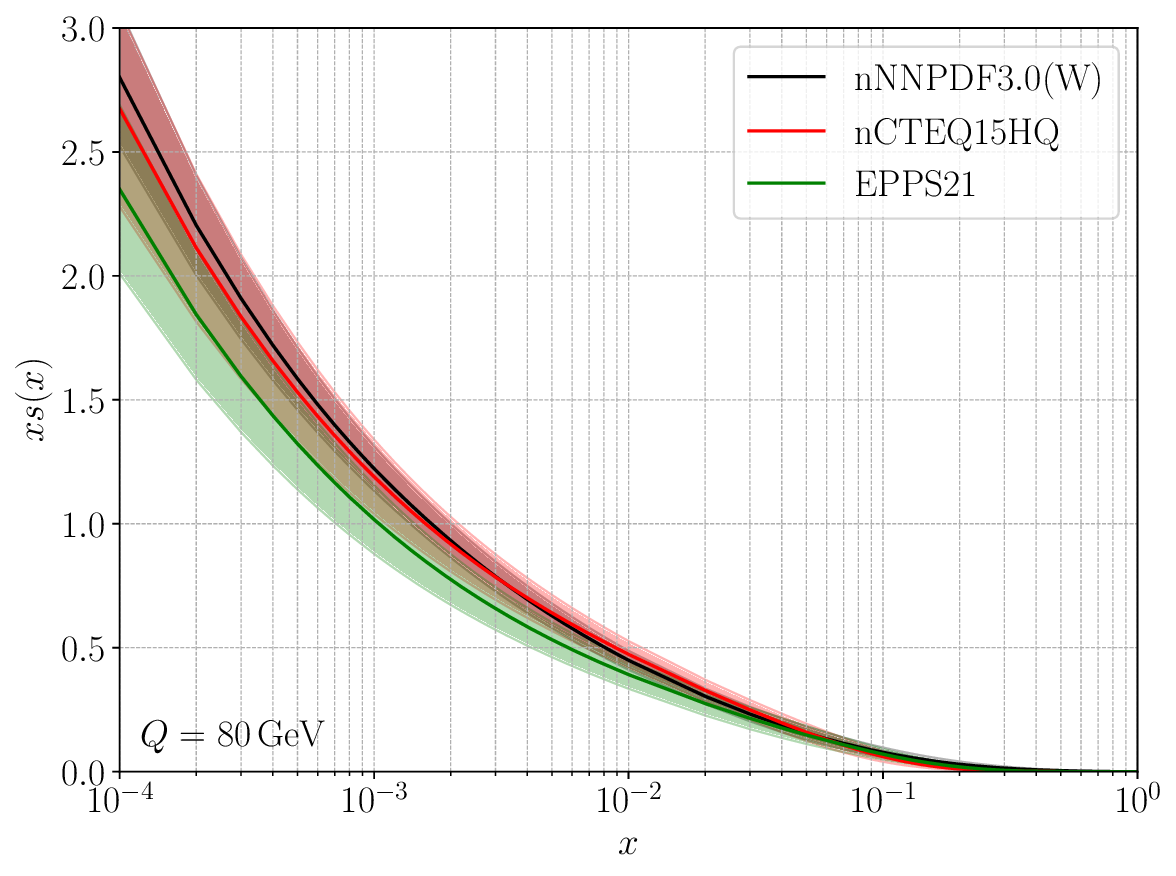}
	\includegraphics[width=0.33\textwidth]{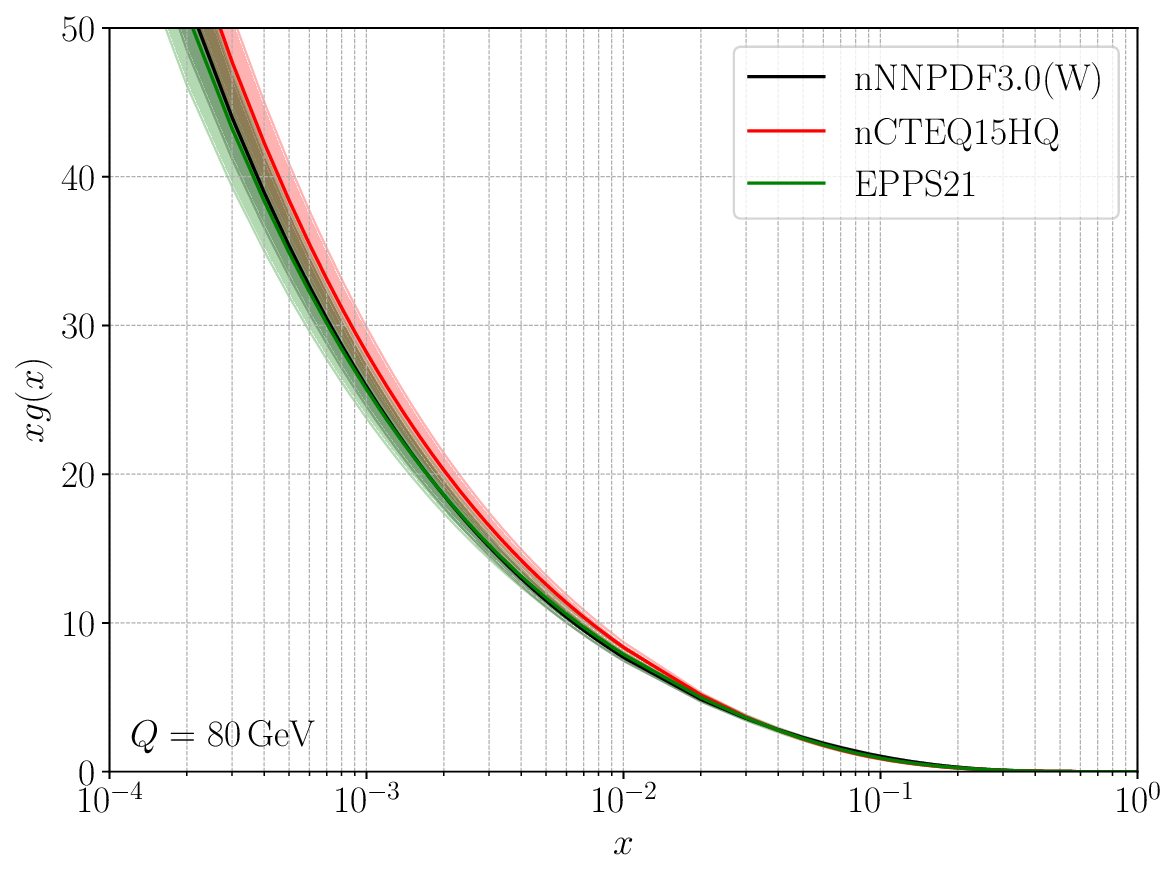}
    
			\end{tabular}
\caption{Comparison between the predictions for the up (left), strange (center) and gluon (right) PDFs of tungsten, derived considering the EPPS21 \cite{Eskola:2021nhw},  nNNPDF 3.0(W) \cite{AbdulKhalek:2022fyi} and nCTEQ15HQ \cite{Duwentaster:2022kpv} parameterizations. Results for two values of the boson virtuality: $Q=2$~GeV (upper panels) and $Q=80$~GeV (lower panels). Uncertainty bands correspond to 68\% confidence levels.  }
\label{fig:pdfs_nuclear}
\end{figure}

One of the main goals of the current hadronic and future electron - ion colliders is the systematic determination of the nuclear parton distribution functions (nPDFs), which are fundamental to improve our understanding of the structure of nuclei and to determine the initial conditions and properties of the Quark Gluon Plasma (QGP) formed in heavy ion collisions.
Over the last decades, different measurements in fixed - target and collider experiments have demonstrated that the nPDFs are not a simple superposition of the free - nucleon parton distributions, with the difference depending on the value of the Bjorken - $x$ variable (For a recent review see, e.g., Ref.~\cite{Klasen:2023uqj}). Early DIS data for the nuclear structure function $F_2^A$, measured in charged - lepton DIS on different nuclear targets,  indicated that $F_2^A \lesssim A F_N$ for momentum fractions $x \lesssim 0.1$ (shadowing region) and
$0.3   \lesssim x \lesssim 0.7$ (EMC region), and that $F_2^A \gtrsim  A F_N$ for $0.1 \lesssim x \lesssim 0.3$ (antishadowing region) and $x  \gtrsim 0.7$ (Fermi motion).  Nuclear effects were also observed in charged current (CC) neutrino DIS on heavy nuclear targets and in the Drell - Yan dilepton production at proton - nucleus fixed target collisions. More recently, the study of nuclear collisions at RHIC and LHC provide a large amount of data for several observables sensitive to the medium effects, which has allowed us to reduce the uncertainty on the nuclear PDFs. However, the spread between the predictions derived by different groups \cite{Eskola:2021nhw,AbdulKhalek:2022fyi,Duwentaster:2022kpv} that perform a global analysis of the current data~\footnote{For a detailed discussion about the distinct methodologies used by these groups, we refer the interested reader to Ref.~\cite{Klasen:2023uqj}.} is still rather significant, as demonstrated in Fig. \ref{fig:pdfs_nuclear}, especially at small values of the hard scale $Q$, low $x$ and in the strange and gluon distributions. 
Another important result that was obtained in these global analyses is that the inclusion of neutrino DIS data diminish the quality of the fit, indicating a tension between different datasets. In particular, Ref.~\cite{Muzakka:2022wey}  
has confirmed that the existing charged lepton - ion DIS data are incompatible with the bulk of the $\nu A$ DIS  data, which has motivated  an intense debate about the breakdown of the universality of the nuclear PDFs some years ago\cite{Schienbein:2007fs,Paukkunen:2010hb,Kovarik:2010uv,Paukkunen:2013grz}. However, the interpretation that this tension is associated with problems in the acquisition of the neutrino - ion data cannot be discarded \cite{Muzakka:2022wey}. In addition, it is important to emphasize that a systematic tension was not observed in the analysis performed by NNPDF group \cite{AbdulKhalek:2022fyi}.
These results demonstrate that new data for neutrino - ion and charged lepton - ion deep inelastic scattering in the GeV -- TeV range is needed to improve our understanding of the nuclear effects and decide about the universality (or not) of nPDFs.

The far-forward physics program of the LHC has recently reported the detection of the first neutrinos from colliders with SND@LHC \cite{SNDLHC:2022ihg,SNDLHC:2023pun} and FASER \cite{FASER:2023zcr,FASER:2024hoe,FASER:2024ref} collaborations. These neutrinos come from the decay of hadrons produced in the forward direction of the interaction point (IP) in the ATLAS detector, being sensitive to different QCD approaches used to simulate the hadroproduction in proton-proton collisions \cite{Kling:2023tgr,John:2025qlm}. It is known that not only neutrinos are produced in the forward direction, but as these far-forward detectors are located approximately 480~m from the ATLAS IP, only neutrinos and muons produced in pp collisions can reach them. As a consequence, the far - forward detectors at the LHC can be used to  measure the deep inelastic scattering cross - sections associated with neutrino - ion \cite{Cruz-Martinez:2023sdv} and muon - ion  \cite{Francener:2025pnr} interactions, as well as to search for beyond the standard model physics \cite{Ariga:2023fjg,MammenAbraham:2025gai}. In this paper, we will investigate the impact of nuclear effects in the muon and neutrino DIS at FASER$\nu$ and its proposed update, FASER$\nu$2, expected to operate during the High-Luminosity (HL) era of the LHC \cite{Anchordoqui:2021ghd,Feng:2022inv}. Our analysis is motivated by the  studies performed in Refs.~\cite{Francener:2025pnr,Cruz-Martinez:2023sdv}, which have shown that muon and neutrino DIS events at FASER$\nu$ will cover a large range in $x$ ($3\times 10^{-4} - 0.9$), i.e., will allow us to investigate the impact of the distinct nuclear effects (shadowing, anti-shadowing, EMC and Fermi motion) on the observables. In addition, as the LHC far-forward program will be able to measure both neutrino and muon DIS using the same detector and nuclear target, it will allow us to perform a direct test of the universality of  nuclear PDFs.
In our analysis, we will consider two classes of events generated in neutrino - tungsten and muon - tungsten interactions: (a) inclusive, and (b) charm tagged events. Our motivation for the selection of these events is associated with the fact that they probe different kinematical ranges of $x$ and are sensitive to distinct parton distributions. In particular, charm tagged events are strongly dependent on the gluon (strange) PDF in muon (neutrino) - ion interactions, which are the distributions with the largest uncertainties, as shown in Fig. \ref{fig:pdfs_nuclear}. Moreover, we will present results for the ratio between charm tagged and inclusive events, which can be useful to discriminate between the distinct parameterizations of the nuclear effects. 


This paper is organized as follows. In the next section, we present a brief review of the formalism and tools used in our calculations, as well as the experimental setup is described. In Section \ref{sec:res}, we present our results for the rates associated with inclusive and charmed tagged events, derived considering different parameterizations for the nPDFs. The impact of the nuclear effects is estimated in muon and neutrino DIS at FASER$\nu$ and FASER$\nu 2$ detectors. Finally, in Section \ref{sec:sum}, we summarize our main results and conclusions.

\section{Lepton - ion DIS at the LHC}
\label{sec:for}
In what follows, we will present a brief review of the  calculation of the number of events associated with lepton - ion DIS at the LHC (For more details see, e.g., Refs.~\cite{Francener:2025pnr,Cruz-Martinez:2023sdv}). 
We will focus on a charged (neutral) current scattering in the case of an incident neutrino (muon). The process is represented by the reaction
\begin{eqnarray}
 l_i + W \rightarrow l_f + X_h \,,   
\end{eqnarray}
where $l_i$ is the incident lepton that scatters into the tungsten target $W$, and the final state is characterized by a charged lepton, $l_f$, and an inclusive hadronic final state, $X_h$.  The number of events expected in the detector can be expressed as \cite{Francener:2025pnr,vanBeekveld:2024ziz}
\begin{eqnarray}
    N_{events} = 
    \int_{0}^{1} \int_{0}^{1} \int_{Q^{2}_{min}}^{Q^{2}_{max}}
    \mathrm{d}x\, \mathrm{d}x_{l_i}\,\mathrm{d}Q^{2}\,  \frac{\mathrm{d}^{2}\sigma_{l_i\, W}}{\mathrm{d}x\, \mathrm{d}Q^{2}}f(x_{l_i})\, \mathcal{A}(E_{l_f}, n_{\mathrm{tr}}) \, ,
    \label{eq:events}
\end{eqnarray}
where $\mathrm{d}\sigma_{l_i\, W}/\mathrm{d}x\,\mathrm{d}Q^{2}$ is the lepton-tungsten differential cross-section, $x$ and $Q^{2}$ the DIS variables, $f(x_{l_i})$ the lepton PDF flux with energy fraction $x_{l_i}$ and $\mathcal{A}$ describes the detector acceptance cuts as a function of the outgoing lepton energy ($E_{l_f}$) and the number of charged tracks in the final state with a minimum momentum ($n_{\mathrm{tr}}$). 

One of the main ingredients for evaluate the events defined in Eq.~(\ref{eq:events}) is the lepton flux $f(x_{l_i})$ that reaches the detector, which contains the information associated with its geometry and time exposure. The muon and anti-muon fluxes in the far-forward direction of the ATLAS were simulated with FLUKA \cite{Sabate-Gilarte:2023aeg,Battistoni:2015epi}, and recently the authors of the Ref. \cite{Francener:2025pnr} have made it available in the LHAPDF format \cite{Buckley:2014ana}. They provided a PDF for the muon flux, $f(x_{\mu})$, with $x_{\mu}$ being the muon energy fraction compared with the proton incident energy at the LHC. A similar analysis for the neutrino flux was performed in Ref. \cite{vanBeekveld:2024ziz}, which also provided a PDF for the neutrino flux in the LHAPDF format, but in this case $x_{\nu} = E_\nu / \sqrt{s_{\mathrm{pp}}} = E_\nu / 2E_p$. This neutrino PDF has been constructed using  the neutrino flux from light mesons obtained in \cite{Kling:2021gos}, while the neutrino flux that comes from heavy meson decay was obtained from \cite{Buonocore:2023kna}.

The lepton-tungsten cross-sections will be estimated  using the POWHEG-BOX-RES event generator \cite{Banfi:2023mhz,FerrarioRavasio:2024kem}, which simulates lepton DIS with fixed target at next-to-leading order (NLO) corrections of perturbative QCD. Partonic cross-sections are then interfaced to PYTHIA8 for hadronization \cite{vanBeekveld:2024ziz,Bierlich:2022pfr}, assuming the Monash 2013 tune \cite{Skands:2014pea}. Moreover, we will assume  the three distinct parameterizations for the nuclear PDFs presented in Fig. \ref{fig:pdfs_nuclear}: 
EPPS21 \cite{Eskola:2021nhw}, nCTEQ15HQ \cite{Duwentaster:2022kpv} and nNNPDF 3.0 \cite{AbdulKhalek:2022fyi}, which will be denoted as nNNPDF 3.0(W) hereafter. In particular, we will use the nNNPDF 3.0 set derived including the LHCb data for the $D$ - meson production at forward rapidities. For comparison, we also will present the predictions derived disregarding the nuclear effects, calculated assuming  using the proton CT18ANLO parameterization \cite{Hou:2019qau} and the baseline for the free nucleon from nNNPDF 3.0 parameterization \cite{AbdulKhalek:2022fyi}, which will be denoted NNPDF 3.0(p) hereafter. Both parameterizations are rescaled for a tungsten target.


The last ingredient in Eq.~(\ref{eq:events}) needed to estimate the number of events is the acceptance of the detector, $\mathcal{A}$. In our analysis will consider the same acceptance used in \cite{Francener:2025pnr}: final lepton energy larger than 100 GeV, and at least two charged tracks in the hadronic final state having momentum larger than 1 GeV each. For semi-inclusive final states, where a charm is produced, we will assume the charm tag efficiency of $\epsilon = 0.7$. In addition to detector acceptance, we will select the DIS events with $Q \geq 1.65\, \mathrm{GeV}$ and  invariant mass of the hadronic final state larger than $2\, \mathrm{GeV}$.

Finally, we will present predictions for the FASER$\nu$ and FASER$\nu 2$ detectors. For FASER$\nu$, we will assume  an integrated luminosity of 250~$\mathrm{fb}^{-1}$, which is expected for FASER$\nu$ data collection during run 3 of the LHC. FASER$\nu$ is operating with 1.1 metric tons of tungsten distributed in 730 layers of 25~cm$\times$30~cm$\times$1.1~mm. In our analysis, we set the detector target length to 50~cm for muon DIS, since the muon identification, as well as the incoming and outgoing muon momentum measurements require a propagation over a few centimeters. It is important to emphasize that FASER$\nu$ detector is also expected to operate during Run 4 with an total integrated luminosity of $650\, \mathrm{fb}^{-1}$. On the other hand, 
FASER$\nu$2 is a detector of the proposed Forward Physics Facility \cite{Anchordoqui:2021ghd,Feng:2022inv}, which is expected to operate during the HL-LHC era. This detector will be an upgrade compared to FASER$\nu$ with $\approx 20$ metric tons of tungsten. In our analysis, we will assume the operation during the HL-LHC era with a time-integrated luminosity of $3\, \mathrm{ab}^{-1}$.

\begin{figure}[t]
	\centering
	\begin{tabular}{ccc}
	\includegraphics[width=0.33\textwidth]{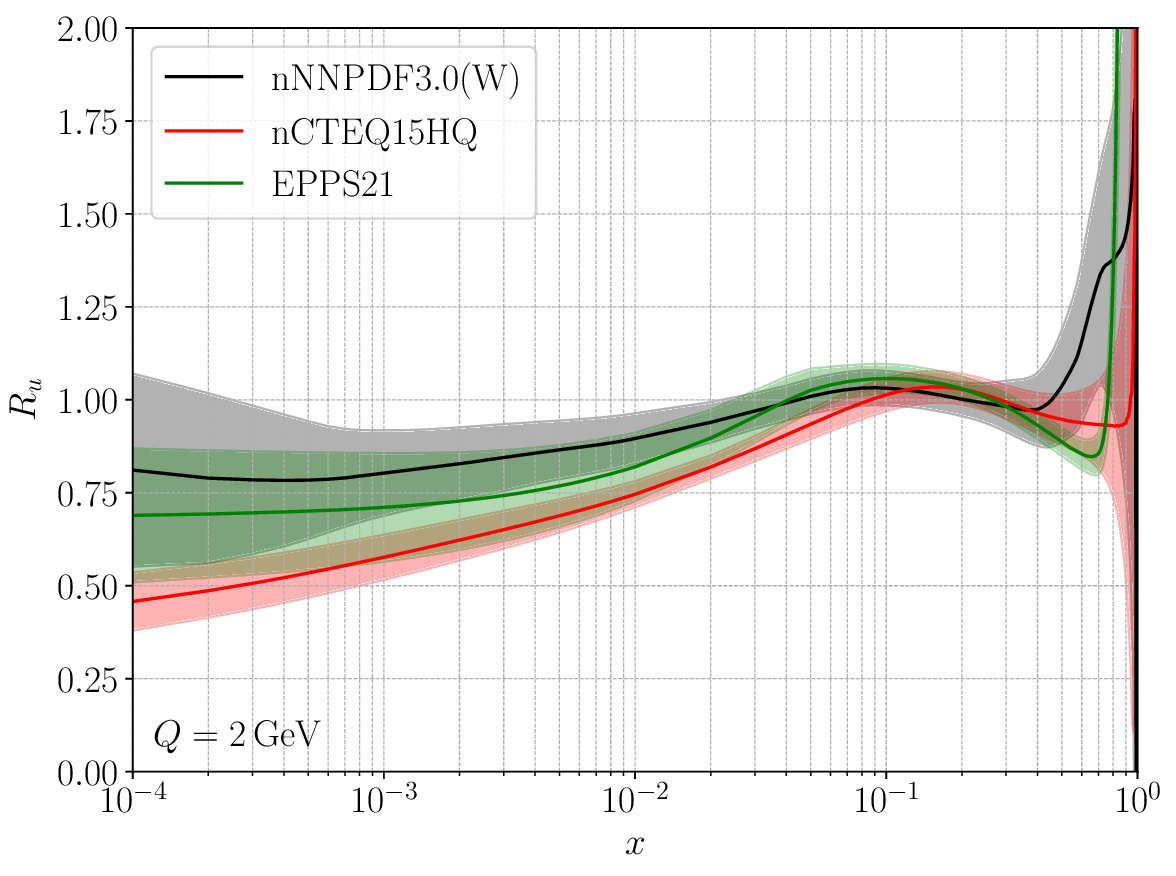}
	\includegraphics[width=0.33\textwidth]{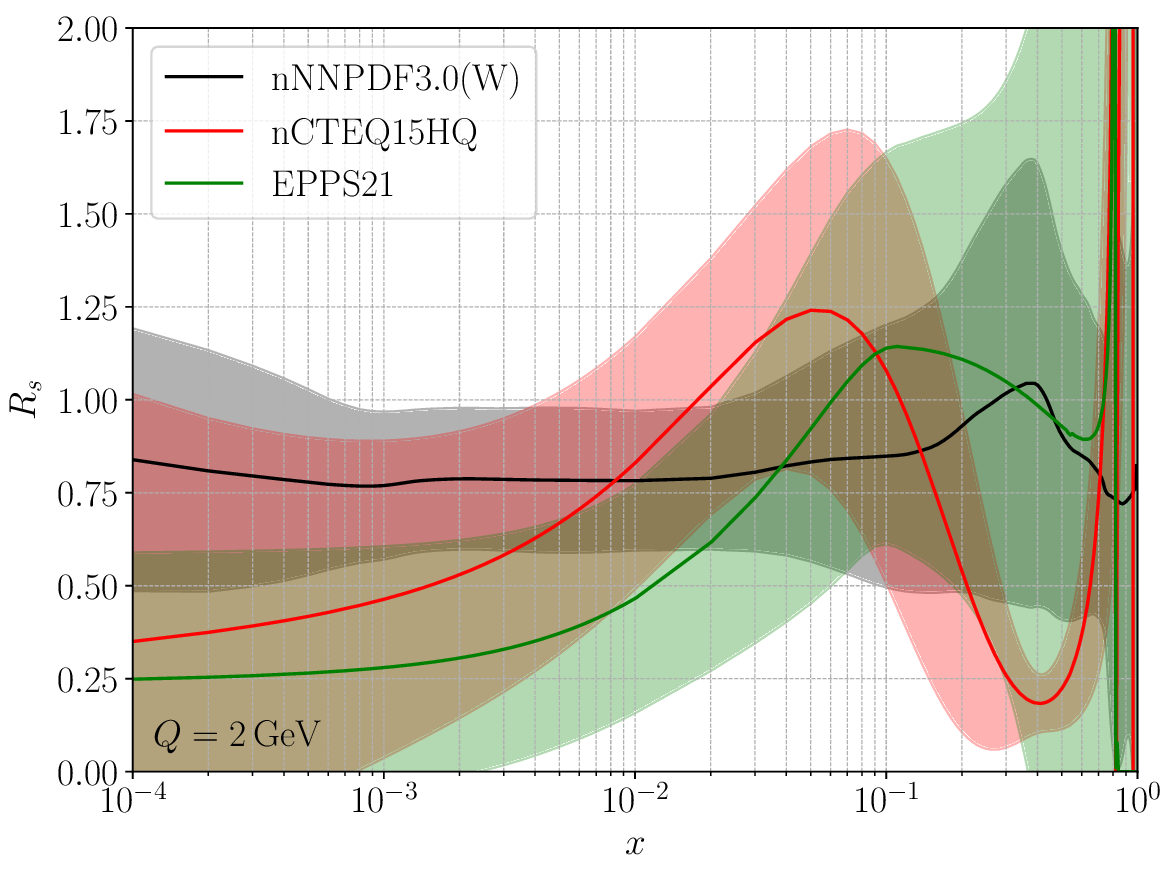}
	\includegraphics[width=0.33\textwidth]{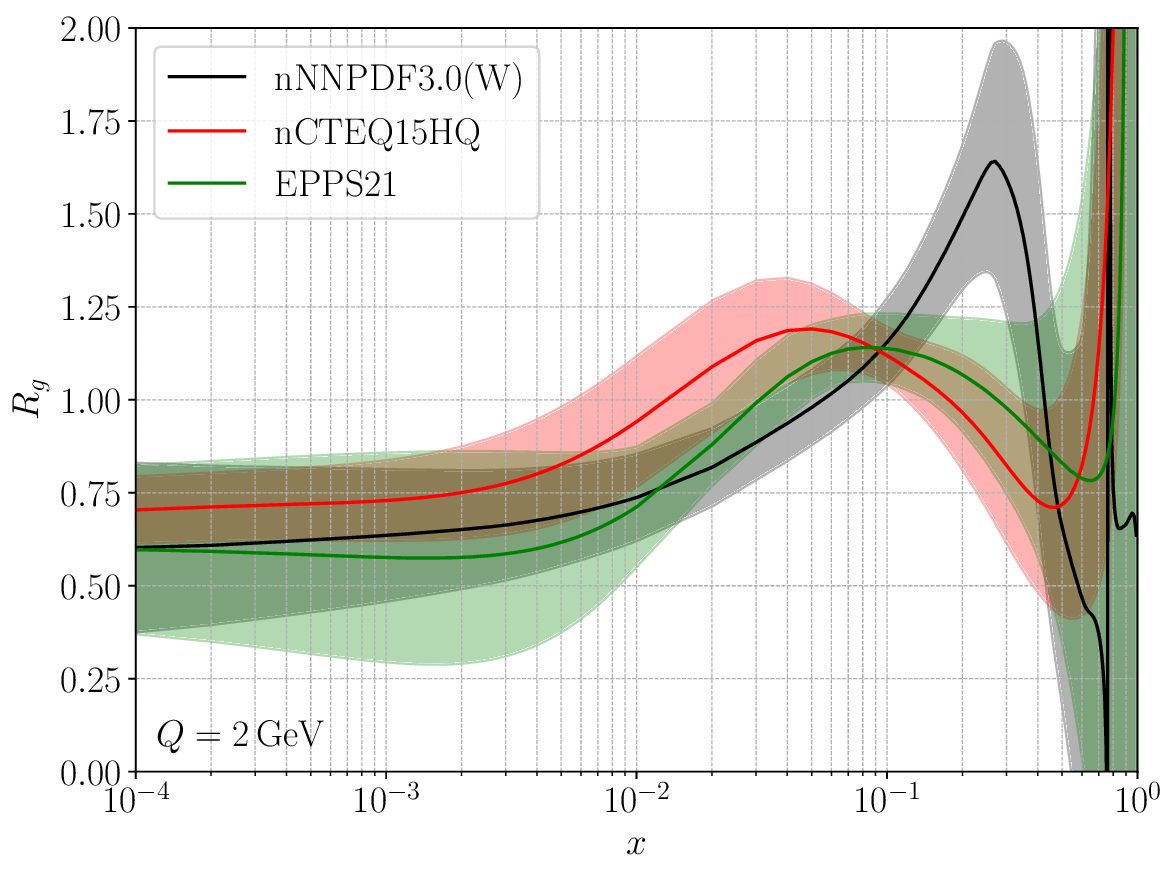}  \\
    \includegraphics[width=0.33\textwidth]{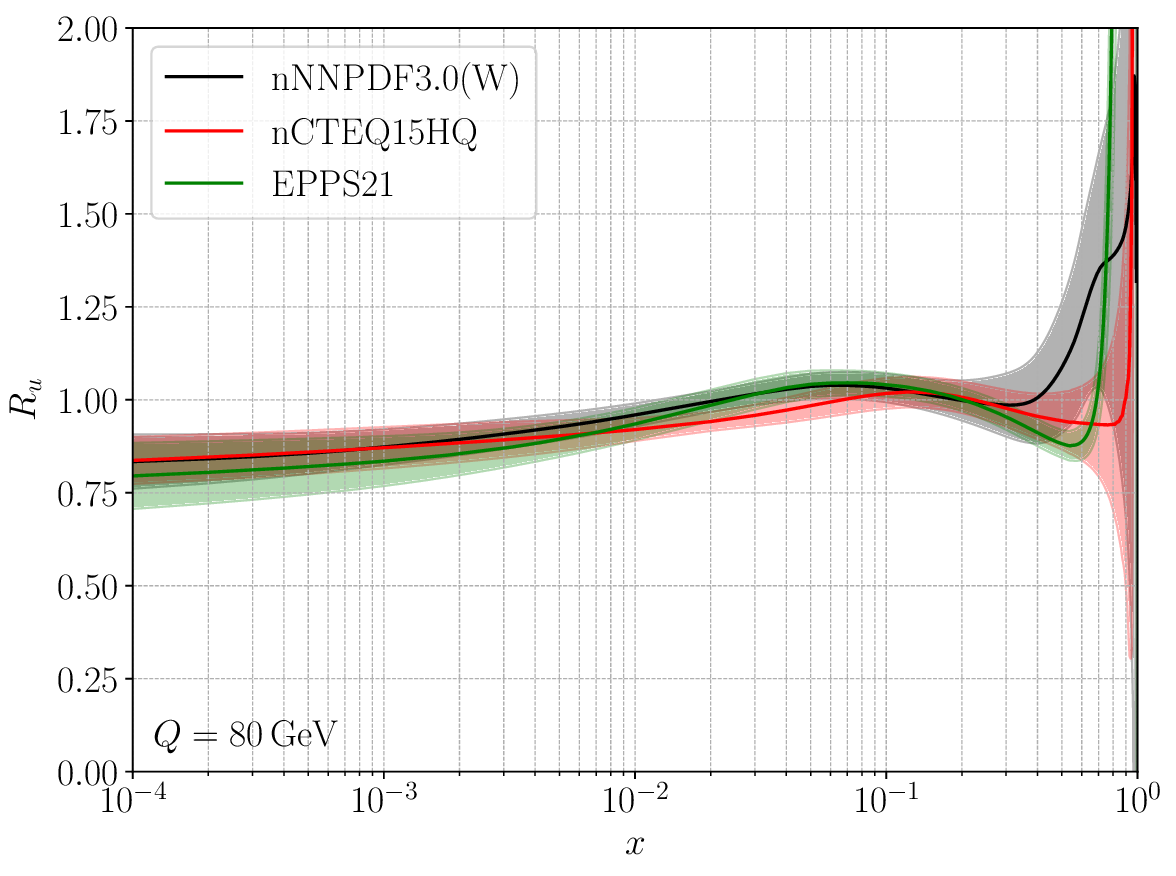}
	\includegraphics[width=0.33\textwidth]{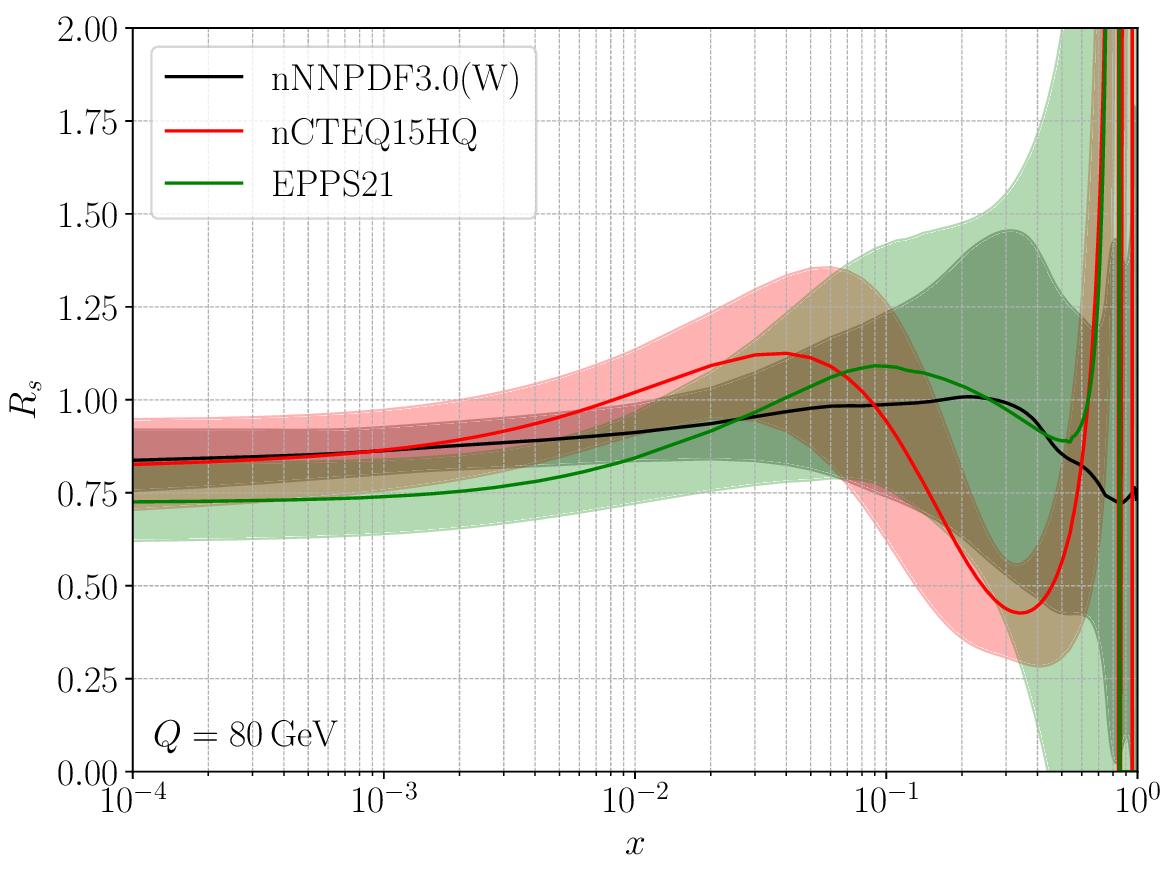}
	\includegraphics[width=0.33\textwidth]{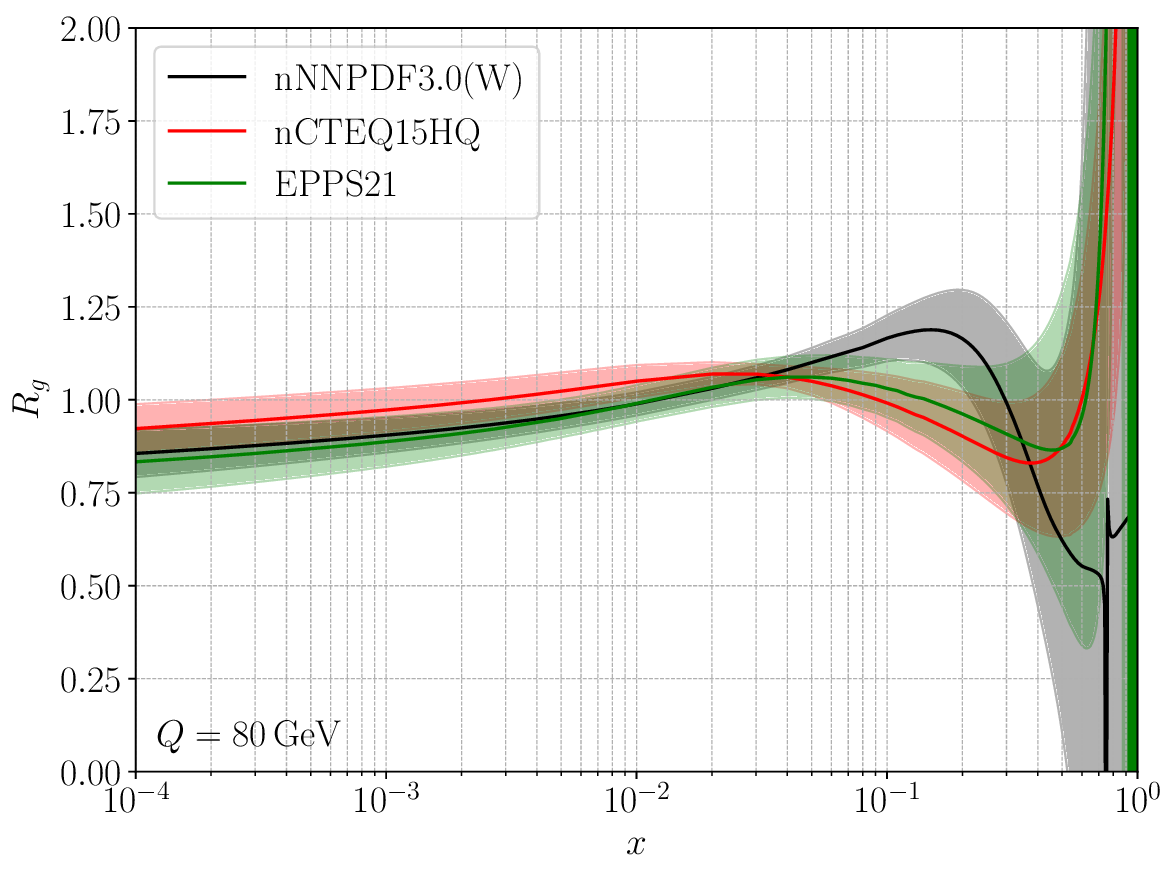}
    
			\end{tabular}
\caption{Ratio between the predictions with and without nuclear effects for the up (left), strange (center) and gluon (right) PDFs of tungsten, derived considering the EPPS21 \cite{Eskola:2021nhw},  nNNPDF 3.0(W) \cite{AbdulKhalek:2022fyi} and nCTEQ15HQ \cite{Duwentaster:2022kpv} parameterizations. Results for two values of the boson virtuality: $Q=2$~GeV (upper panels) and $Q=80$~GeV (lower panels). Uncertainty bands correspond to 68\% confidence levels.  }
\label{fig:ratios_pdfs_nuclear}
\end{figure}

\section{Results}
\label{sec:res}

One of the main goals of this study is to investigate the potential of leptons produced in $pp$ collisions at the LHC and measured by far-forward detectors to improve our understanding of nuclear effects in DIS. In what follows, we will present our results for the events rates considering the muon - tungsten and neutrino - tungsten interactions at the FASER$\nu$ and FASER$\nu 2$ detectors.
Motivated by the recent FASER analyzes \cite{FASER:2024hoe,FASER:2024ref}, which have measured the cross - sections differential in neutrino - energy and rapidity, we also will present predictions for the number of events binned in the Bjorken - $x$ variable. Such a distribution allow us to estimate the impact of the nuclear effects on the cross - sections at different kinematical ranges.

Initially,  in order to illustrate the magnitude of the nuclear effects in the up, strange and gluon distributions, predicted by the distinct parameterizations used in our calculations, we present in Fig. \ref{fig:ratios_pdfs_nuclear} the results for the nuclear ratio $R_i = f_i^A / (A \times f_i^p)$ ($i = u, \,s,\,g$)  considering two values for the boson virtuality $Q$. We have that the impact of the nuclear effects decreases at larger values of $Q$. In addition, for the up quark (left panels), we have that the predictions are similar for $x \approx 0.1$, but become distinct for smaller values of $x$, with the nCTEQ15HQ (nNNPDF 3.0(W)) parameterization predicting the larger (smaller) amount of shadowing. In contrast, for the gluon case(right panels), the central predictions are similar at small - $x$, but largely differ in the amount and position for the maximum of the antishadowing. Finally, for the case of the strange distribution (center panels), we have that the predictions of the distinct parameterizations are very distinct, especially for low values of $Q$.


\begin{table}[t]
\centering
\renewcommand{\arraystretch}{1.5}
\begin{tabularx}{\textwidth}{XXcc}
\toprule
\hline
\multicolumn{3}{c}{\bf Lepton DIS events -- Inclusive case} \\
\hline
\midrule
{\bf PDF set}                         & {\bf Process}		                                & {\bf Number of events}\\
\hline	
\toprule   
\multirow{2}{*}{CT18ANLO}       &  $\mu^\pm + W \rightarrow \mu^\pm + X$        & 1.75$\times10^{5}$ (2.60$\times10^{7}$)         \\
                                &  $\nu_l + W \rightarrow l^\pm + X$            & 5.66$\times10^{3}$ (7.67$\times10^{5}$)         \\
\hline
\midrule
\multirow{2}{*}{nCTEQ15HQ}      &  $\mu^\pm + W \rightarrow \mu^\pm + X$        & 1.54$\times10^{5}$ (2.29$\times10^{7}$)         \\
                                &  $\nu_l + W \rightarrow l^\pm + X$            & 5.66$\times10^{3}$ (7.68$\times10^{5}$)         \\
\hline
\midrule
\multirow{2}{*}{EPPS21}         &  $\mu^\pm + W \rightarrow \mu^\pm + X$        & 1.60$\times10^{5}$ (2.38$\times10^{7}$)         \\
                                &  $\nu_l + W \rightarrow l^\pm + X$            & 5.55$\times10^{3}$ (7.52$\times10^{5}$)         \\
\hline   
\midrule
\multirow{2}{*}{nNNPDF3.0(p)}   &  $\mu^\pm + W \rightarrow \mu^\pm + X$        & 1.77$\times10^{5}$ (2.64$\times10^{7}$)         \\
                                &  $\nu_l + W \rightarrow l^\pm + X$            & 5.81$\times10^{3}$ (7.88$\times10^{5}$)         \\
\hline
\midrule
\multirow{2}{*}{nNNPDF3.0(W)}   &  $\mu^\pm + W \rightarrow \mu^\pm + X$        & 1.71$\times10^{5}$ (2.55$\times10^{7}$)         \\
                                &  $\nu_l + W \rightarrow l^\pm + X$            & 6.12$\times10^{3}$ (8.29$\times10^{5}$)         \\
\hline
\bottomrule
\end{tabularx}
\vspace{0.3cm}
\caption{Predictions for the number of inclusive events in $\mu W$ and $\nu W$ interactions at FASER$\nu$ (FASER$\nu$2), derived considering different parameterizations for the nPDFs and assuming an integrated luminosity of $\mathcal{L}_{\rm pp}=250$ fb$^{-1}$ (3 ab$^{-1}$).
}
\label{table:Nevents}
\end{table}

\begin{table}[t]
\centering
\renewcommand{\arraystretch}{1.5}
\begin{tabularx}{\textwidth}{XXcc}
\toprule
\hline
\multicolumn{3}{c}{\bf Lepton DIS events -- Charm tagged case}\\
\hline
\midrule
{\bf PDF set}                         & {\bf Process}		                                & {\bf Number of events}\\
\hline	
\toprule   
\multirow{2}{*}{CT18ANLO}       &  $\mu^\pm + W \rightarrow \mu^\pm + X + c(\bar{c})$        & 9.85$\times10^{3}$ (1.36$\times10^{6}$)         \\
                                &  $\nu_l + W \rightarrow l^\pm + X + c(\bar{c})$            & 281 (3.82$\times10^{4}$)                        \\
\hline
\midrule
\multirow{2}{*}{nCTEQ15HQ}      &  $\mu^\pm + W \rightarrow \mu^\pm + X + c(\bar{c})$        & 9.91$\times10^{3}$ (1.47$\times10^{6}$)         \\
                                &  $\nu_l + W \rightarrow l^\pm + X + c(\bar{c})$            & 263 (3.56$\times10^{4}$)                        \\
\hline
\midrule
\multirow{2}{*}{EPPS21}         &  $\mu^\pm + W \rightarrow \mu^\pm + X + c(\bar{c})$        & 7.37$\times10^{3}$ (1.39$\times10^{6}$)         \\
                                &  $\nu_l + W \rightarrow l^\pm + X + c(\bar{c})$            & 263 (3.57$\times10^{4}$)                        \\
\hline   
\midrule
\multirow{2}{*}{nNNPDF3.0(p)}   &  $\mu^\pm + W \rightarrow \mu^\pm + X + c(\bar{c})$        & 6.41$\times10^{3}$ (9.53$\times10^{5}$)         \\
                                &  $\nu_l + W \rightarrow l^\pm + X + c(\bar{c})$            & 411 (5.58$\times10^{4}$)                        \\
\hline
\midrule
\multirow{2}{*}{nNNPDF3.0(W)}   &  $\mu^\pm + W \rightarrow \mu^\pm + X + c(\bar{c})$        & 7.14$\times10^{3}$ (1.06$\times10^{6}$)         \\
                                &  $\nu_l + W \rightarrow l^\pm + X + c(\bar{c})$            & 381 (5.13$\times10^{4}$)                        \\
\hline
\bottomrule
\end{tabularx}
\vspace{0.3cm}
\caption{Predictions for the number of charm tagged events in $\mu W$ and $\nu W$ interactions at FASER$\nu$ (FASER$\nu$2), derived considering different parameterizations for the nPDFs and assuming an integrated luminosity of $\mathcal{L}_{\rm pp}=250$ fb$^{-1}$ (3 ab$^{-1}$).
}
\label{table:Nevents_charm}
\end{table}

In Tables \ref{table:Nevents} and \ref{table:Nevents_charm} we present our predictions for the number of events for the inclusive and charm tagged cases, respectively, derived considering muon - tungsten and neutrino - tungsten interactions at the FASER$\nu$ and FASER$\nu2$ detectors. The results for the 
FASER$\nu2$ detector are in parentheses. We have that number of events associated with $\nu W$ interactions are almost two order of magnitude smaller in comparison with the $\mu W$ case. Moreover, as expected from the analyzes performed in Refs.~\cite{Francener:2025pnr,Cruz-Martinez:2023sdv}, the number of events at FASER$\nu 2$ will be a factor $\approx 100$ larger than at FASER$\nu$. Another important aspect is the large number of charm tagged events in $\mu W$ interactions, which are expected to be sensitive to the gluon and charm distributions in the target. For the inclusive case, we have that the CT18ANLO (nNNPDF 3.0(W)) parameterization provides the higher value for the number of events in $\mu W$ ($\nu W$) interactions. In contrast, for charm tagged events, the larger values for the number of events in $\mu W$ and $\nu W$ interactions are generated by the nCTEQ15HQ and nNNPDF3.0(p) parameterizations, respectively. Such results demonstrate the dependence of the predictions on the nPDF parameterization considered in the calculations and motivate a more detailed analysis of less inclusive observables. In the next two subsections, we will present our predictions for the differential distribution binned in $x$.

\begin{figure}[t]
	\centering
	\begin{tabular}{ccc}
	\includegraphics[width=0.5\textwidth]{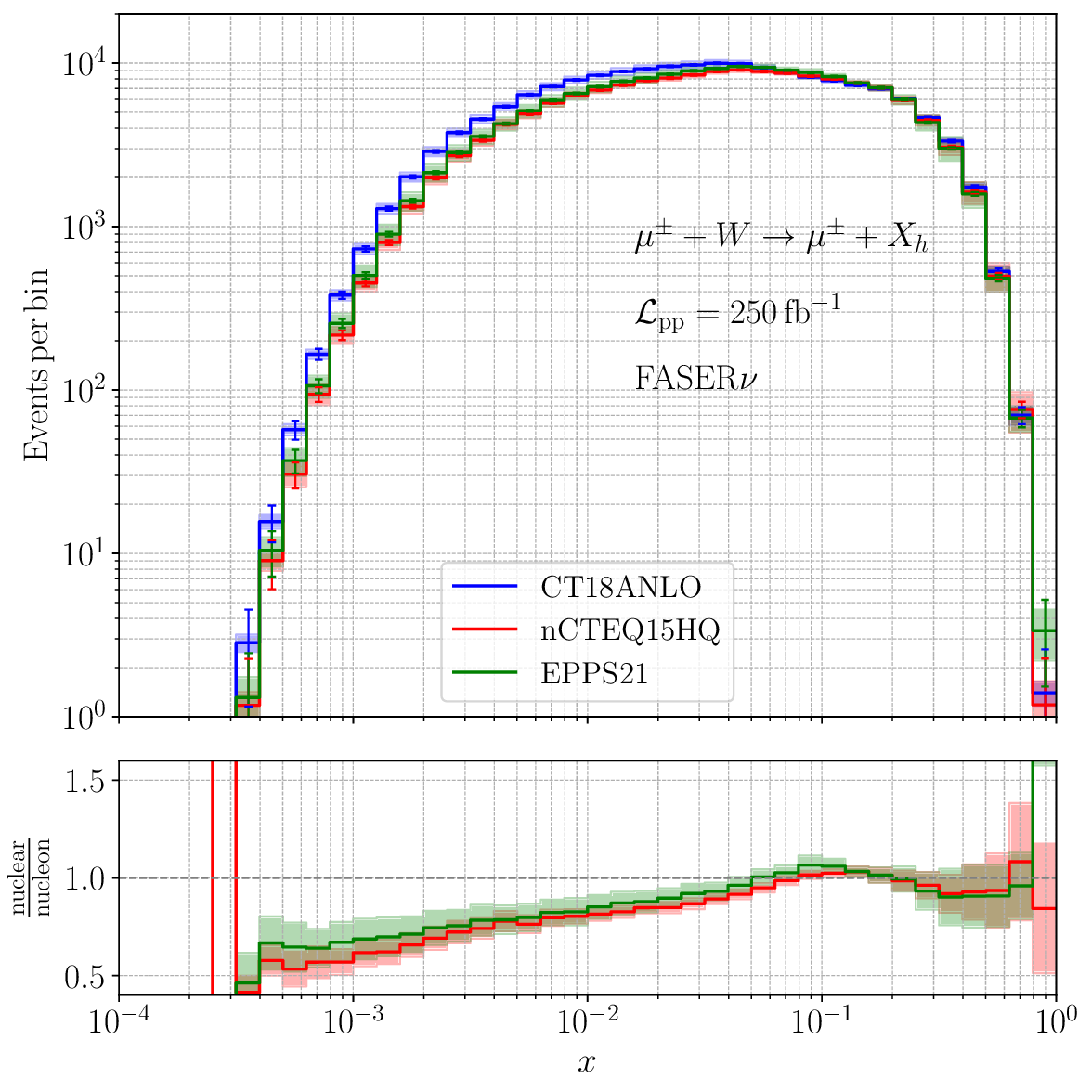}
	\includegraphics[width=0.5\textwidth]{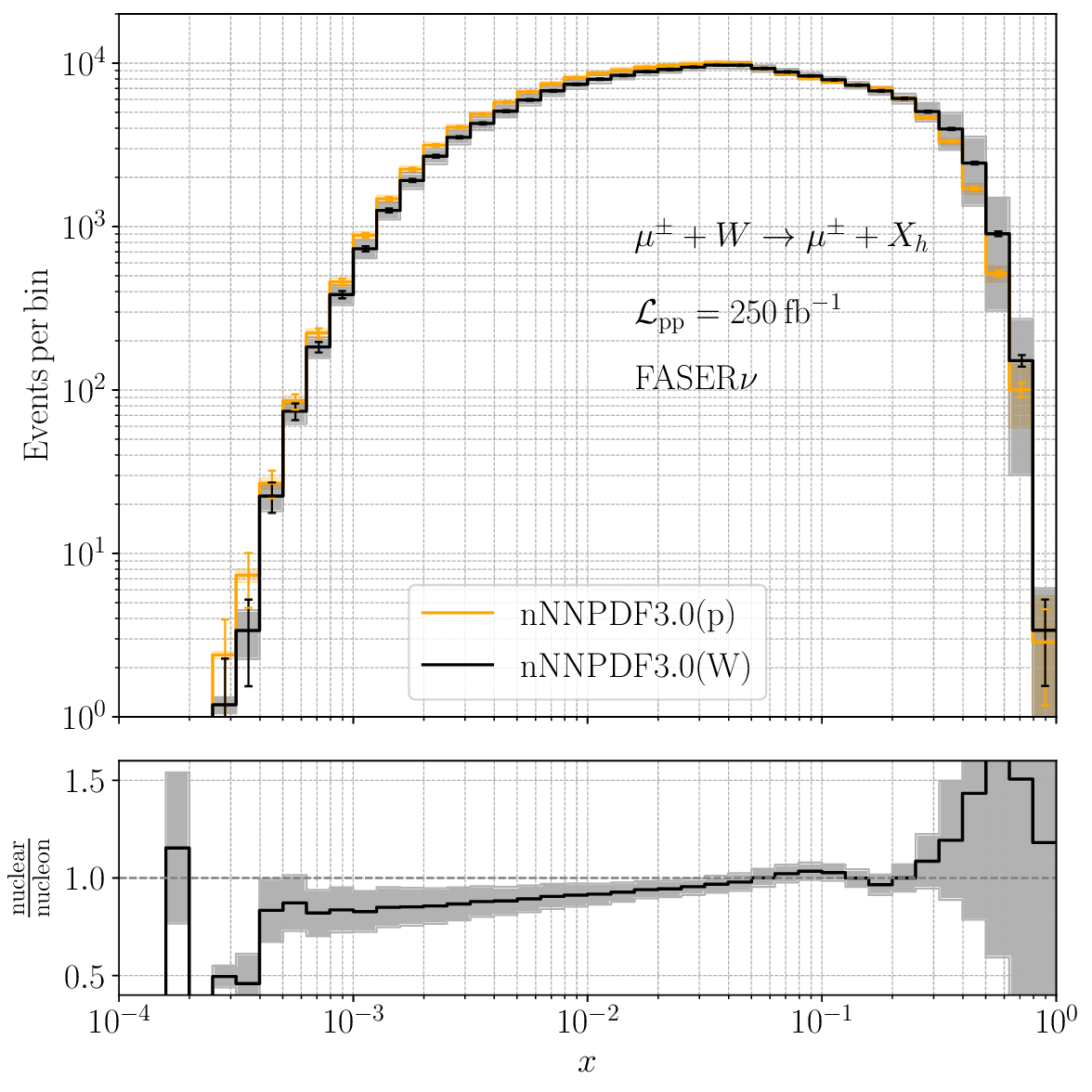} \\
    \includegraphics[width=0.5\textwidth]{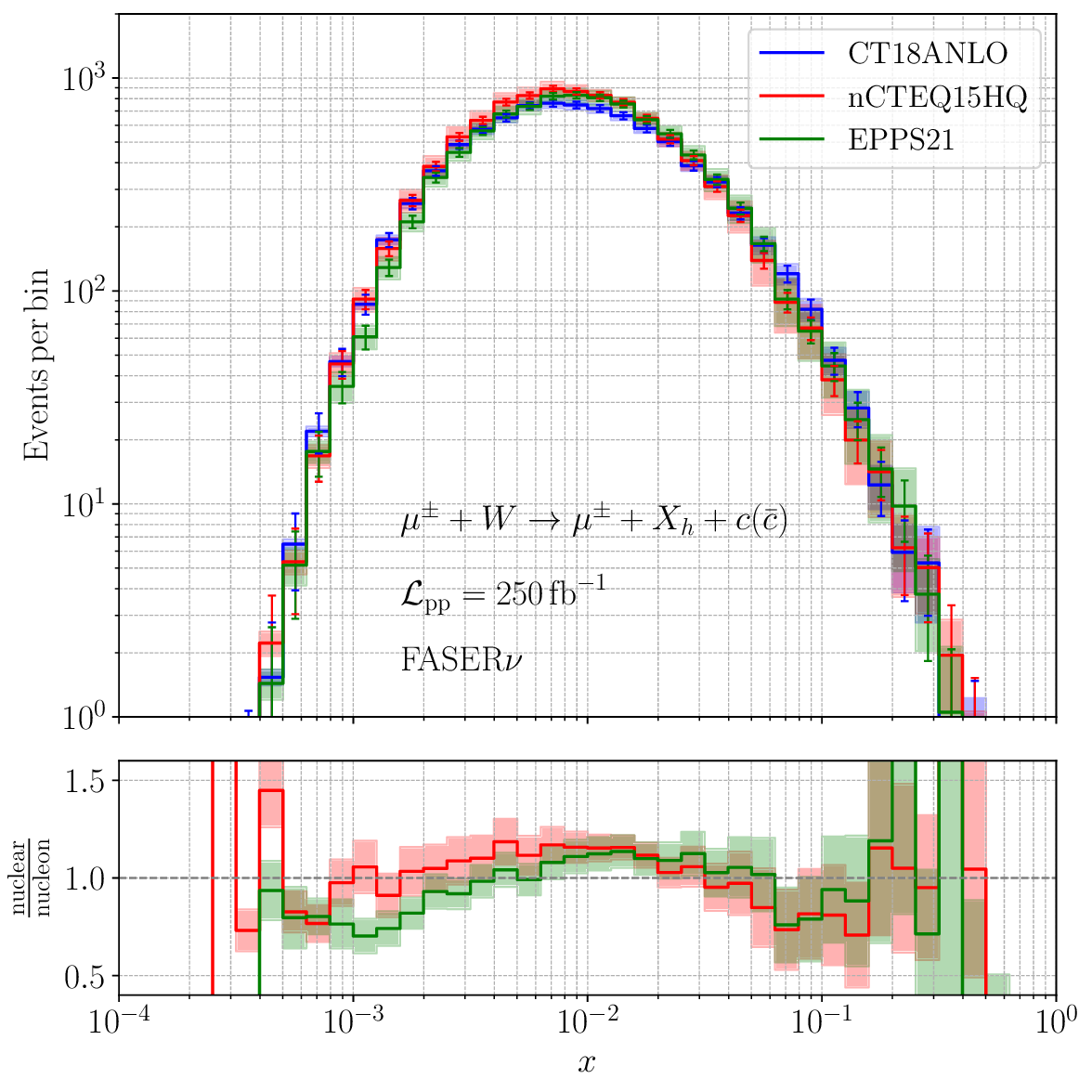}
	\includegraphics[width=0.5\textwidth]{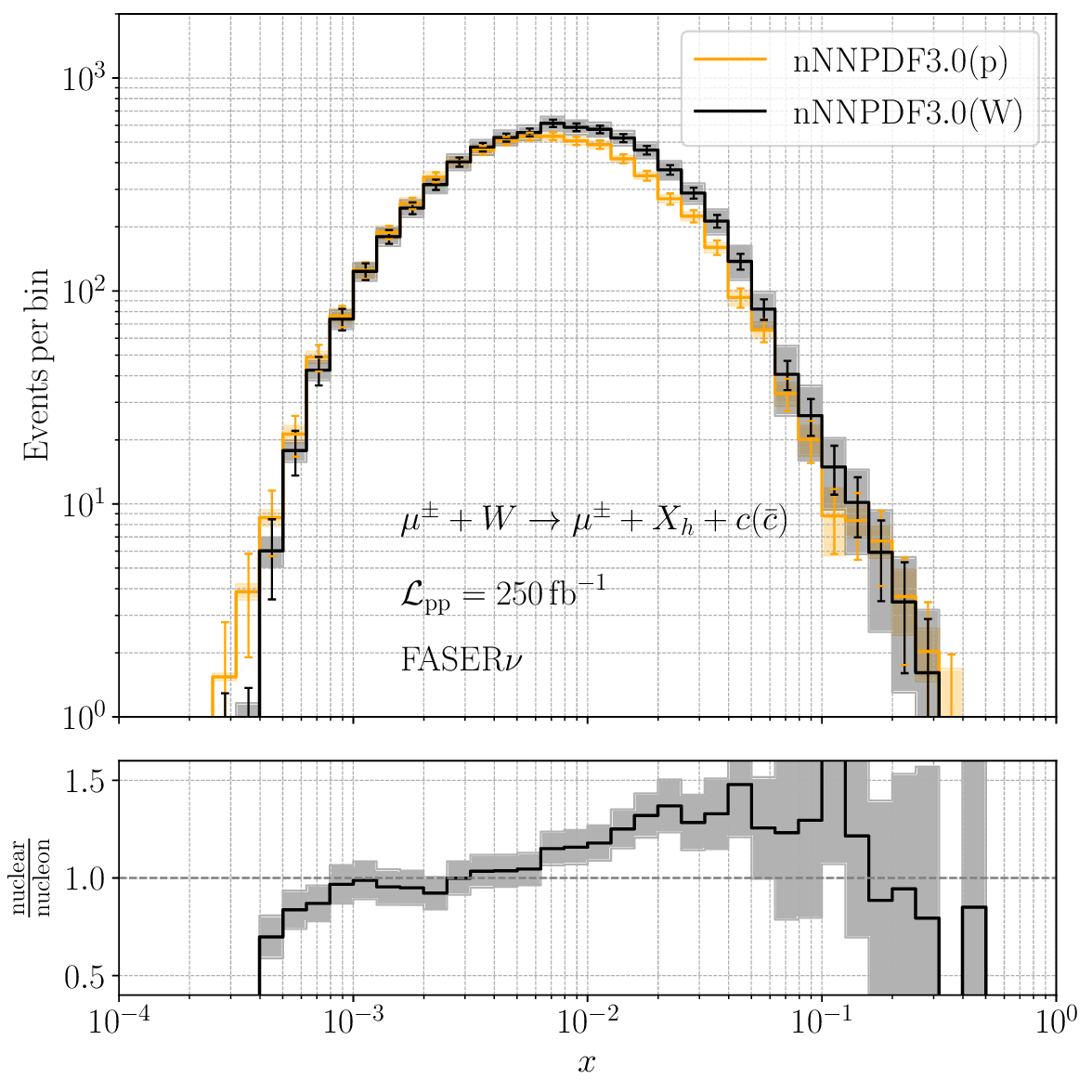}
	\end{tabular}
\caption{ Predictions for the number of events for muon DIS at FASER$\nu$ binned in $x$ in the inclusive case (upper panels) and with  a tagged charm (lower panels) in the final state. Results derived assuming different parameterizations for the nPDFs. The predictions for the ratio between the nuclear and nucleon results are presented in the bottom panels of the plots.}
\label{fig:muonDIS_x_FASERnu}
\end{figure}

\subsection{FASER$\nu$}


In Fig.~\ref{fig:muonDIS_x_FASERnu} we present our predictions for events of muon plus anti-muon DIS binned in $x$ for FASER$\nu$ detector. In the upper panel, we present our predictions for inclusive events, while in the lower panel we have the results for events with at least one charm hadron tagged in the final state. The ratio between the predictions derived with and without nuclear effects is also presented at the bottom of each panel. As in our analysis we are considering PDFs based on different frameworks, we will present two plots for each class of events. 
In the left panels,  we present the predictions based on the CTEQ framework, i.e. the results associated with CT18ANLO parameterization, which disregards the nuclear effects, and the nCTEQ15HQ and EPPS21 predictions, which are obtained modifying the CT18ANLO parameterization to include nuclear effects. On the other hand, in the right panels, we present the results for the nNNPDF framework, where we compare the nNNPDF 3.0(W) predictions with those derived using its baseline for a free nucleon (nNNPDF 3.0(p)). The uncertainty bands come from PDF uncertainties at 68\% confidence level, while the error bar is the expected statistical uncertainty, also at 68\% confidence level, constructed considering Gaussian statistics. The events presented here are after the acceptance cuts discussed in the previous section.

The results presented in the upper panels of Fig.~\ref{fig:muonDIS_x_FASERnu} show that predictions with and without nuclear effects can be distinguished depending on the region in $x$ and the compared models. For inclusive events and at small-$x$, where we expect the shadowing effect, we have a large difference between CT18ANLO and nCTEQ15HQ/EPPS21 predictions. In contrast, this difference does not occur for nNNPDF 3.0, where the impact of shadowing is smaller. At large-$x$, we have a small difference between the predictions for the expected number of events with and without nuclear effects caused by EMC effect, with similar magnitudes when comparing nCTEQ15HQ and EPPS21 predictions. The amount of anti-shadowing for $x \approx 0.1$ is weakly dependent  on the nPDF considered in the calculation, with the position of the anti-shadowing peak being similar for these two parametrizations. In the region $x\gtrsim 0.3$, we observe a larger difference between central predictions from CTEQ and NNPDF frameworks, arising from the distinct behaviors for the up and gluon distributions shown in Fig. \ref{fig:ratios_pdfs_nuclear}. However, the corresponding PDF uncertainty bands remain overlapping.

In the lower panels of Fig.~\ref{fig:muonDIS_x_FASERnu}, we present our results for events with a charm tagged in the final state. As discussed in the Introduction, these events are generated, at leading order, by photon - gluon interactions and are strongly sensitive to the nuclear gluon distribution. We have that in comparison with the inclusive case, the number of events is reduced by one order of magnitude. Moreover, the maximum of events occurs for a smaller value of $x$. We have that the EPPS21 parameterization predicts a smaller number of events at small - $x$ in comparison with the nCTEQ15HQ result. In contrast, these two parameterizations imply a similar number of events for $x \gtrsim 10^{-2}$. On the other hand, for the NNPDF case (lower right panel), the impact of the nuclear effects at small - $x$ is small, but the number of events is enhanced by these effects for $x$ in the range $10^{-2} - 10^{-1}$. 
We also find a large difference between the predictions obtained using the CTEQ and NNPDF frameworks, both with and without nuclear effects. This difference is $\approx 2$ for $x$ in the $10^{-2} - 10^{-1}$ range, which is associated with the distinct charm PDFs provided by these two collaborations. 
This difference originates from the distinct treatments of the perturbative charm contribution adopted in these two frameworks.
At larger values of $x$, where the contribution of a non-perturbative (intrinsic) charm component becomes important,  the predictions of charm tagged events are weakly affected by nuclear effects. 
Although the differences between the central predictions are small,  the uncertainties associated with the nuclear PDFs are large in the  large-$x$ region, with variations of the order of 50\%. However, it is important to emphasize that the non-perturbative (intrinsic) component in the charm PDF increase its PDF in this region by a factor of approximately 10 \cite{Francener:2025pnr}, in addition to giving it a valence-like shape.  
As a consequence, the main conclusions derived disregarding the nuclear effects in Ref. \cite{Francener:2025pnr}  remain valid when these effects are taken into account. 


\begin{figure}[t]
	\centering
	\begin{tabular}{ccc}
	\includegraphics[width=0.5\textwidth]{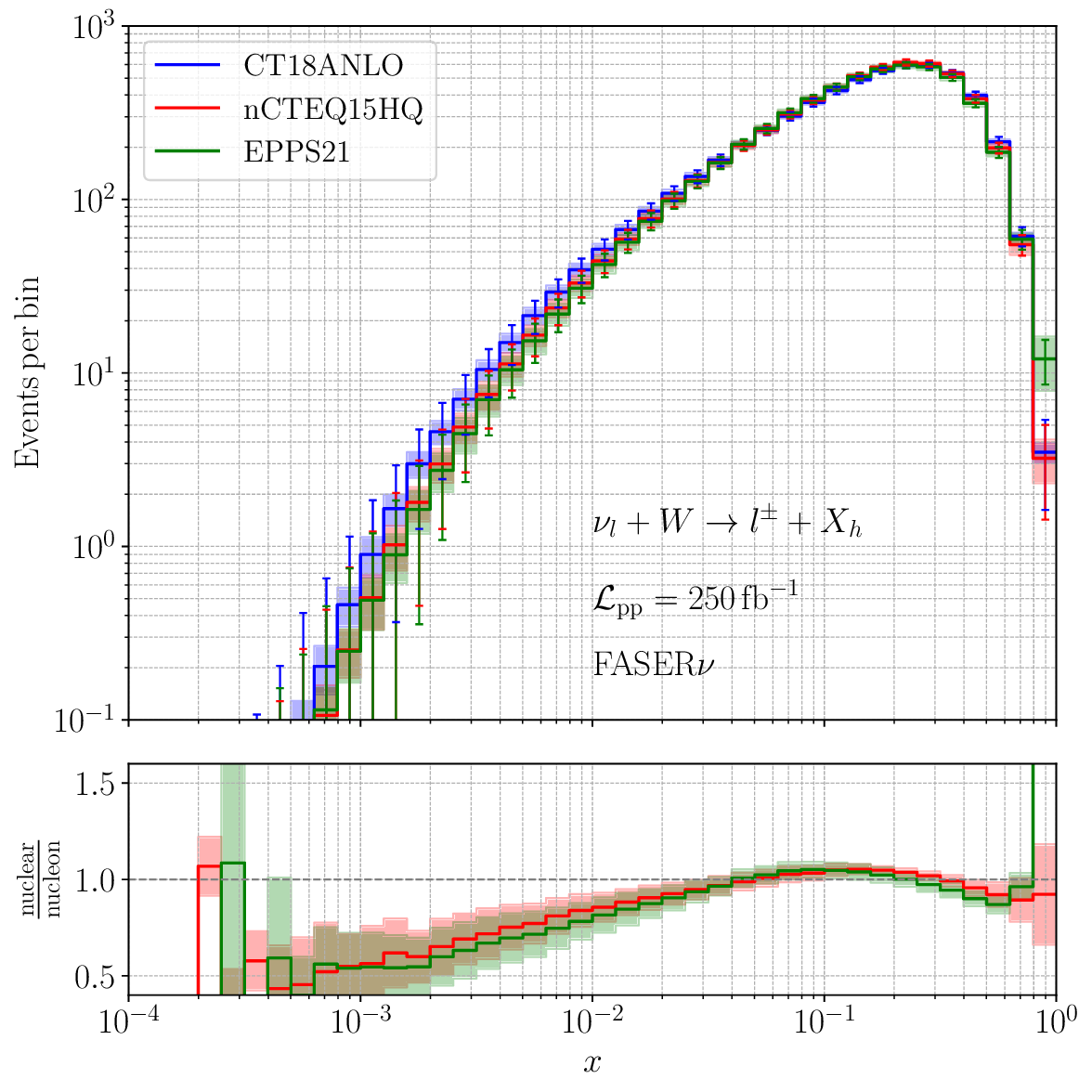}
	\includegraphics[width=0.5\textwidth]{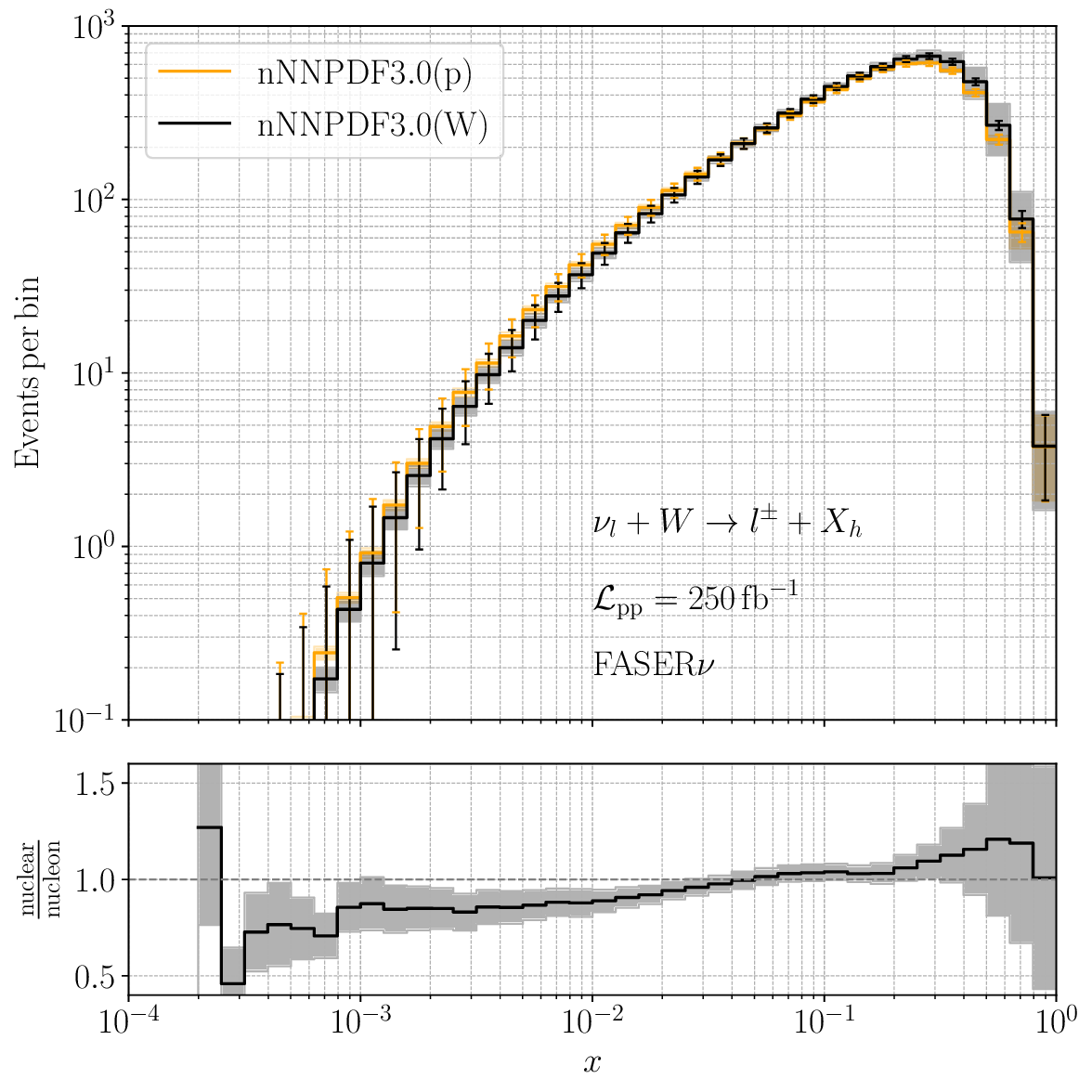} \\
    \includegraphics[width=0.5\textwidth]{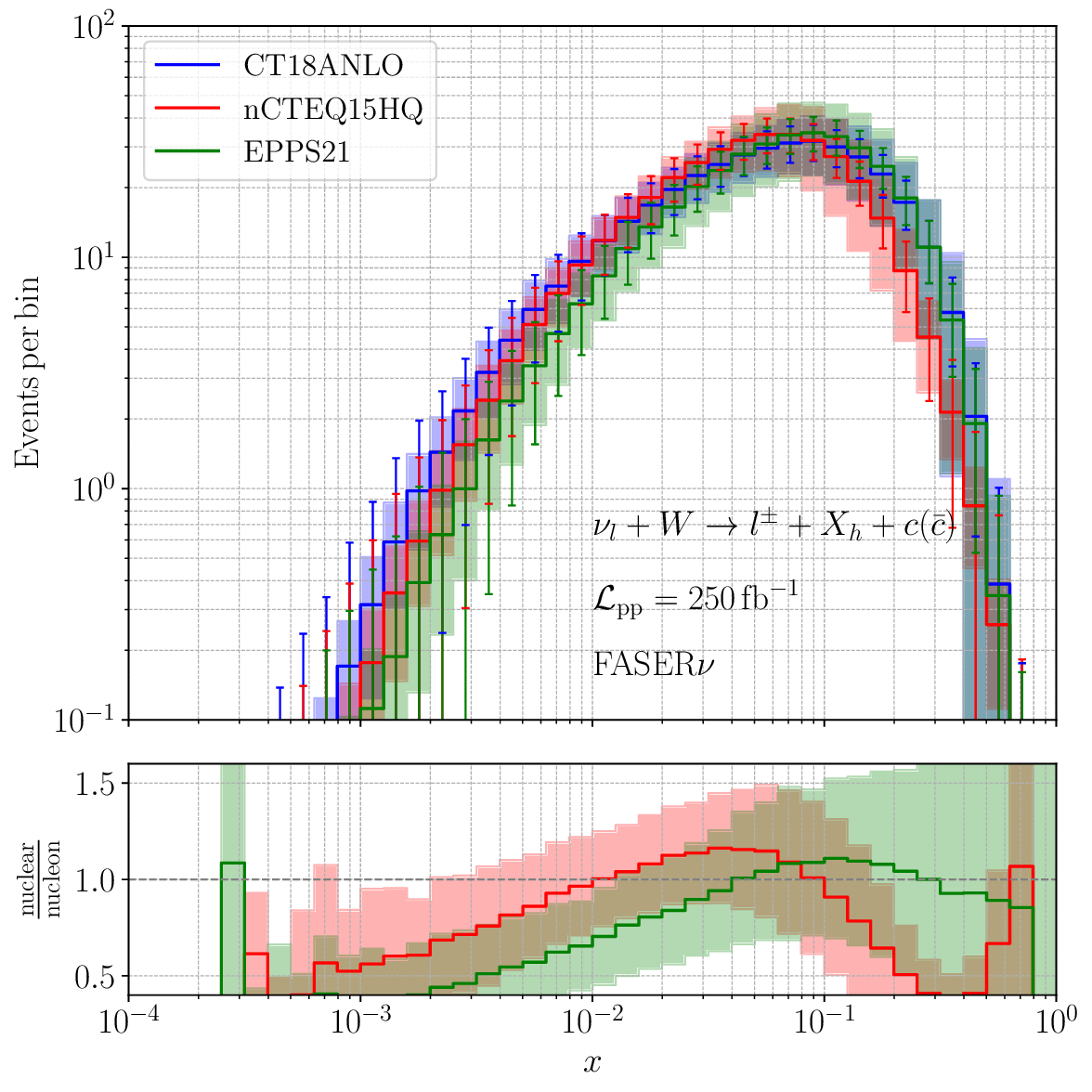}
	\includegraphics[width=0.5\textwidth]{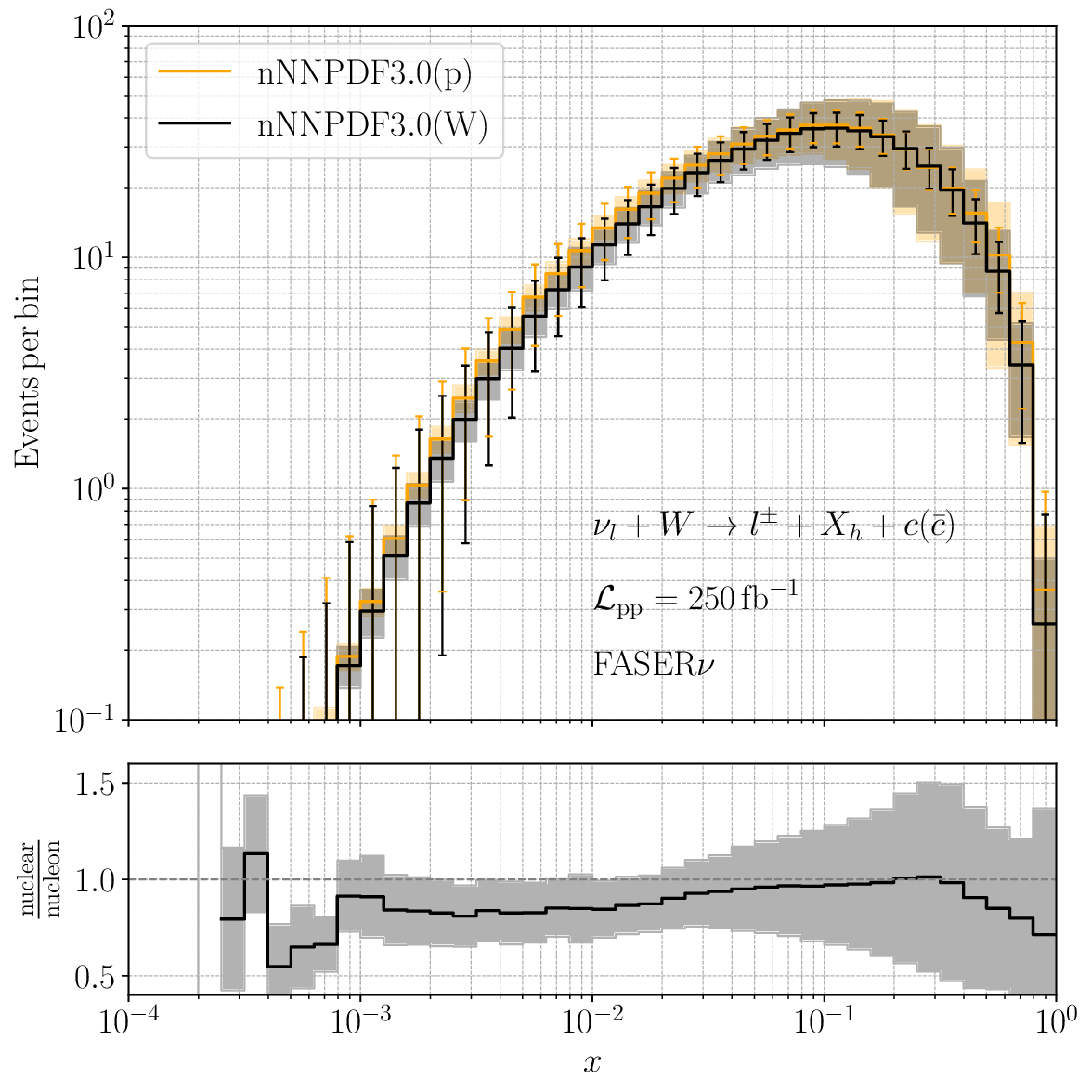}
    
			\end{tabular}
\caption{Predictions for the number of events for neutrino DIS at FASER$\nu$ binned in $x$ in the inclusive case (upper panels) and with  a tagged charm (lower panels) in the final state. Results derived assuming different parameterizations for the nPDFs. The predictions for the ratio between the nuclear and nucleon results are presented in the bottom panels of the plots. }
\label{fig:neutrinoDIS_x_FASERnu}
\end{figure}

Let's now present our predictions for the number of events associated with charged current neutrino - tungsten interactions, which consider the sum over neutrino and antineutrino fluxes as well as over electron and muon flavors. In the upper panels of Fig.~\ref{fig:neutrinoDIS_x_FASERnu}, we present the results for the inclusive case. We have that the expected number of events per bin is smaller when compared with muon-induced events. Moreover, the maximum of events occurs at larger values of $x$, which is associated with the fact that  neutrino DIS are dominated by larger values of $Q^2$. Such a dominance also implies a smaller impact of the nuclear effects for $x \gtrsim 10^{-2}$. As a consequence, the analysis of such events can be useful to constrain the description of the nucleon parton distributions. The results for charm tagged events (lower panels  of Fig.~\ref{fig:neutrinoDIS_x_FASERnu}), indicate that the predictions have a large uncertainty, which is mainly associated with the current uncertainty on the strange PDF (see Figs. \ref{fig:pdfs_nuclear} and \ref{fig:ratios_pdfs_nuclear}). Moreover, we predict few events per bin ($\mathcal{O}(30)$ for $x \approx 0.1$).

\begin{figure}[t]
	\centering
	\begin{tabular}{ccc}
	\includegraphics[width=0.5\textwidth]{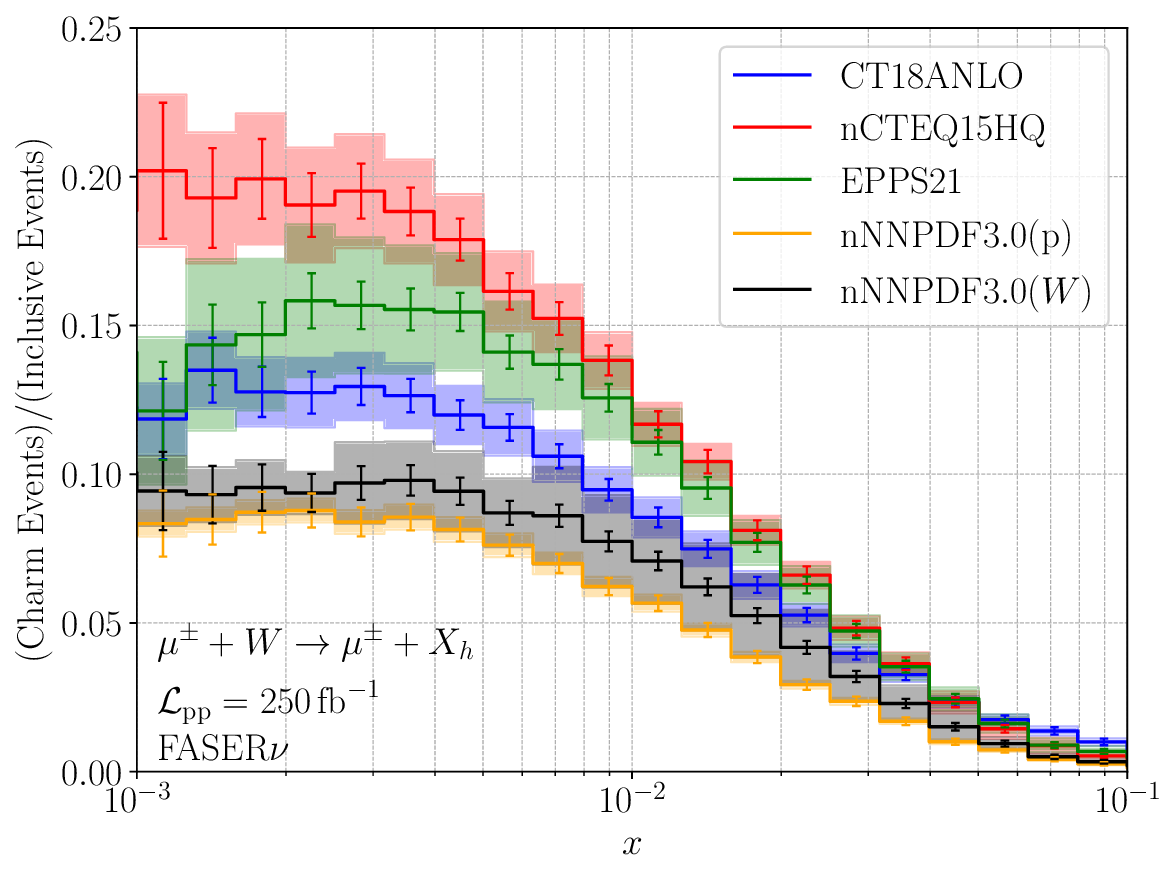}
    \includegraphics[width=0.5\textwidth]{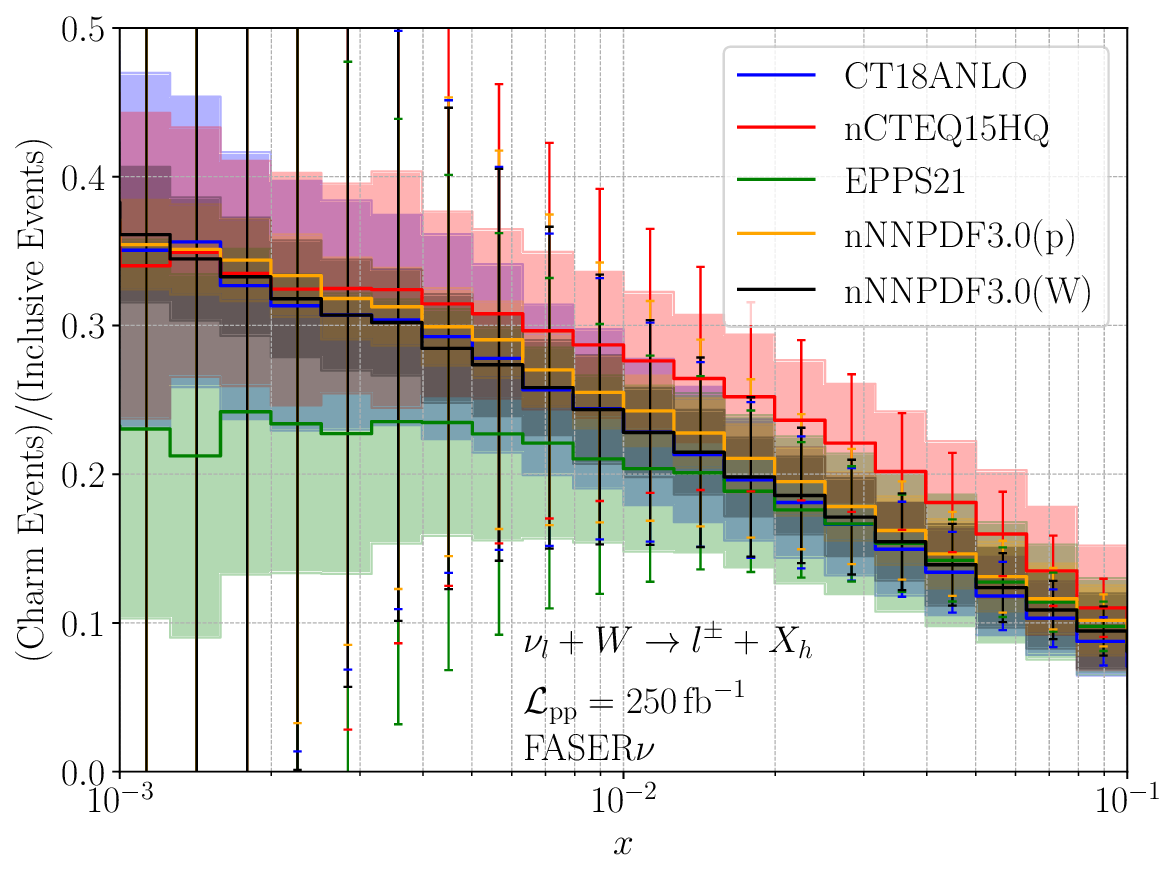}

			\end{tabular}
\caption{ Ratio between events with tagged charm hadron and inclusive for muon (left) and neutrino (right) DIS at FASER$\nu$. Results derived assuming different parameterizations for the nPDFs.}
\label{fig:Ratio_charm_total_FASERnu}
\end{figure}

Considering that the parameterizations predict distinct amounts of nuclear effects in the different parton distributions (See Fig. \ref{fig:ratios_pdfs_nuclear}), an alternative to discriminate between these models is to consider the ratio between cross - sections that have its behaviors determined by distinct PDFs. Here, we propose the analysis of the ratio between the rates for charm tagged and inclusive events. We have that the inclusive events are mainly determined by valence and sea quarks, while charm tagged events are sensitive to the gluon (strange) distribution in the case of muon (neutrino) - induced interactions. Our results for the ratio are presented in Fig. \ref{fig:Ratio_charm_total_FASERnu}. We have that for $\mu W$ DIS events (left panel), the magnitude of the ratio for $x \lesssim 10^{-2}$ is dependent on the nPDF used as input in the calculation, with the nCTEQ15HQ predicting the larger value. In contrast, the results derived using the NNPDF framework are smaller by a factor $\approx 2$.  Another important aspect is that we expect a small statistical  uncertainty in the predictions, which are represented by the vertical lines in the distinct curves.  Such results indicate that a future experimental analysis of this ratio can be useful to improve our understanding of nuclear effects at small - $x$. 
In addition, the analysis of this ratio provides a promising tool to constrain the charm PDF, which exhibits significant differences in the small - $x$ region when comparing the CTEQ and NNPDF parameterizations. 
On the other hand, in the results for $\nu W$ DIS (right panel), we have that the value of the ratio increases by a factor $\approx 2$, with  the EPPS21 (nCTEQ15HQ) parameterization predicting the smaller (larger) value for $x \lesssim 10^{-2}$ ($x \gtrsim 10^{-2}$. However, in this case, we expect  a large  statistical uncertainty at FASER$\nu$, which implies that a stronger conclusion about the more adequate description of the nuclear effects is not possible.


\begin{figure}[t]
	\centering
	\begin{tabular}{ccc}
	\includegraphics[width=0.5\textwidth]{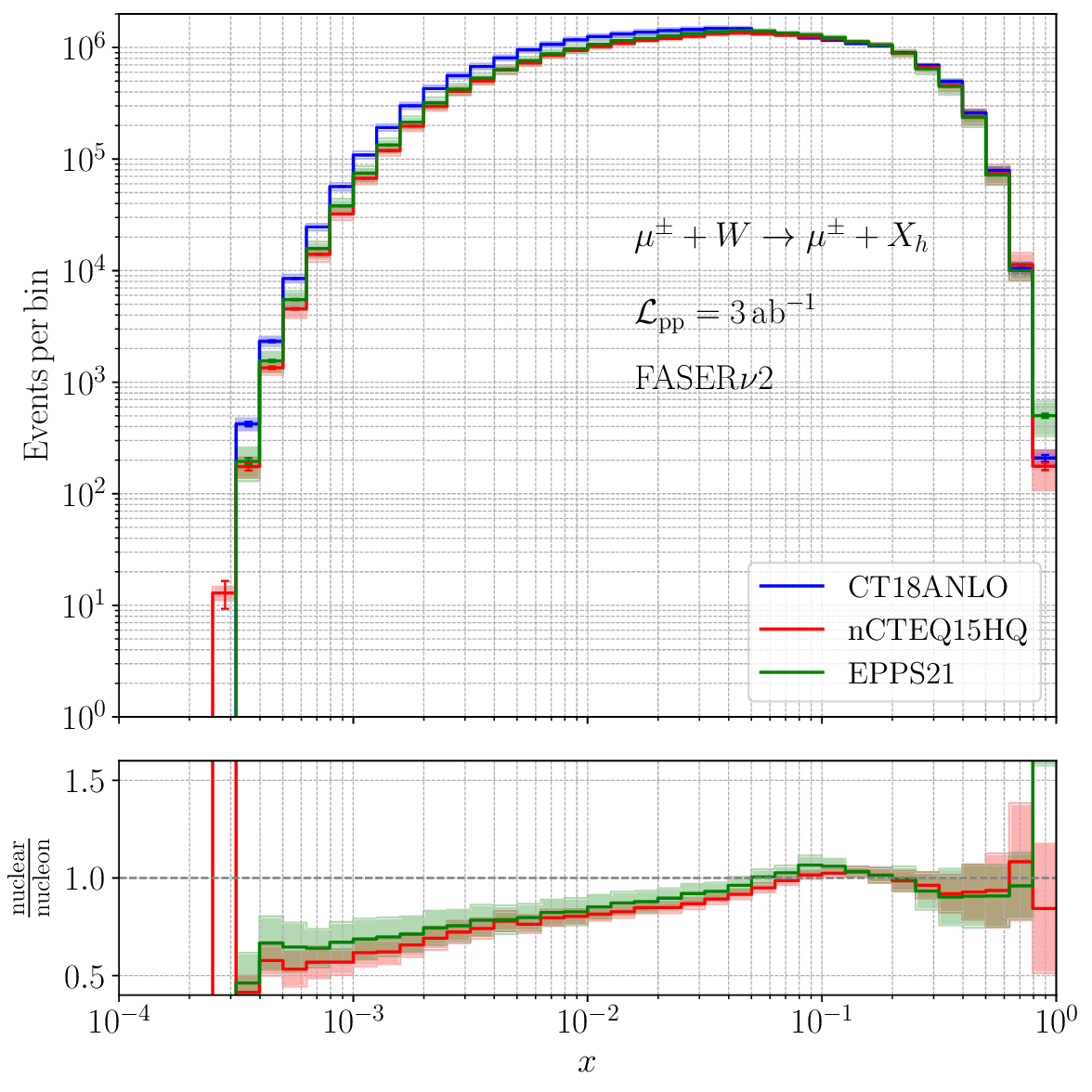}
	\includegraphics[width=0.5\textwidth]{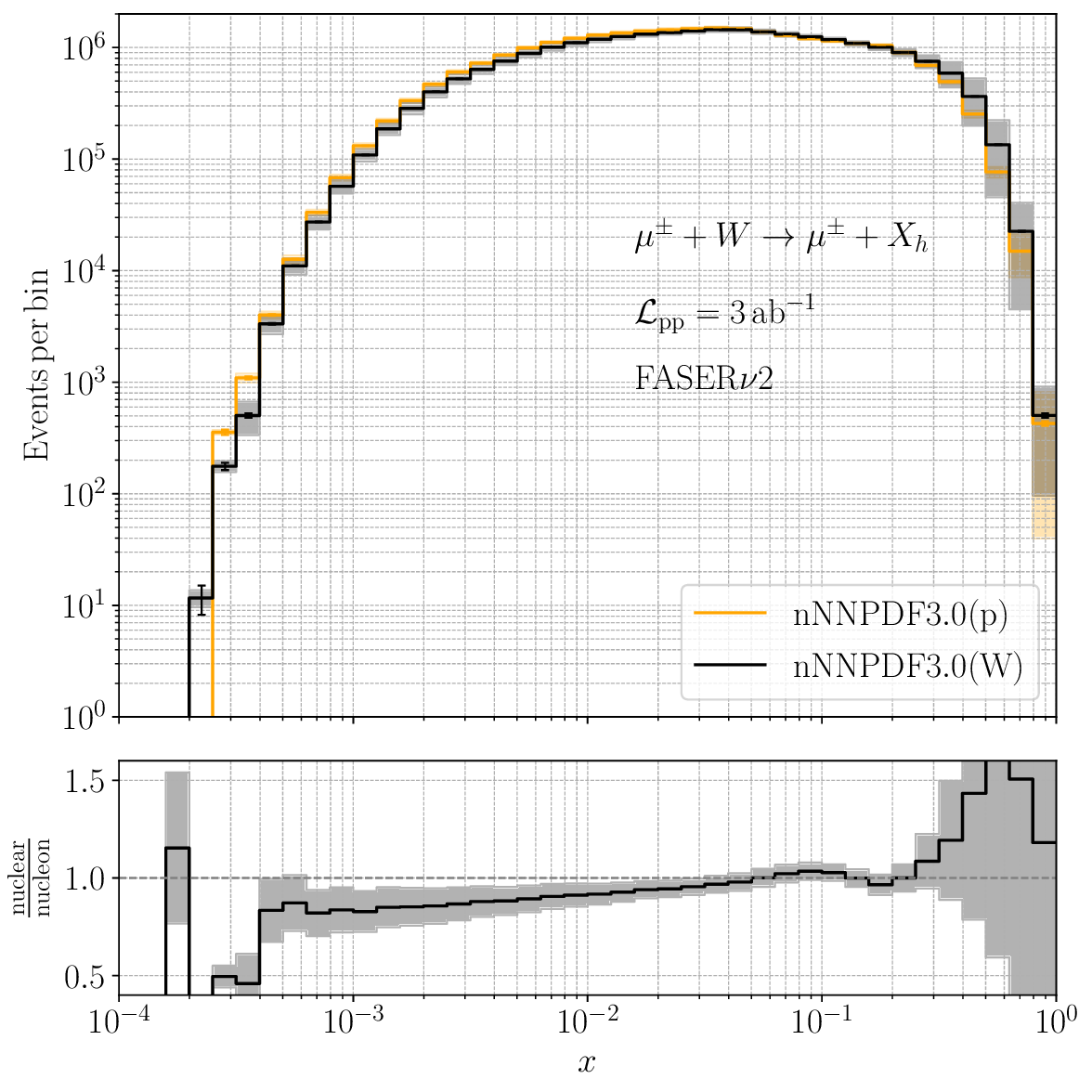} \\
    \includegraphics[width=0.5\textwidth]{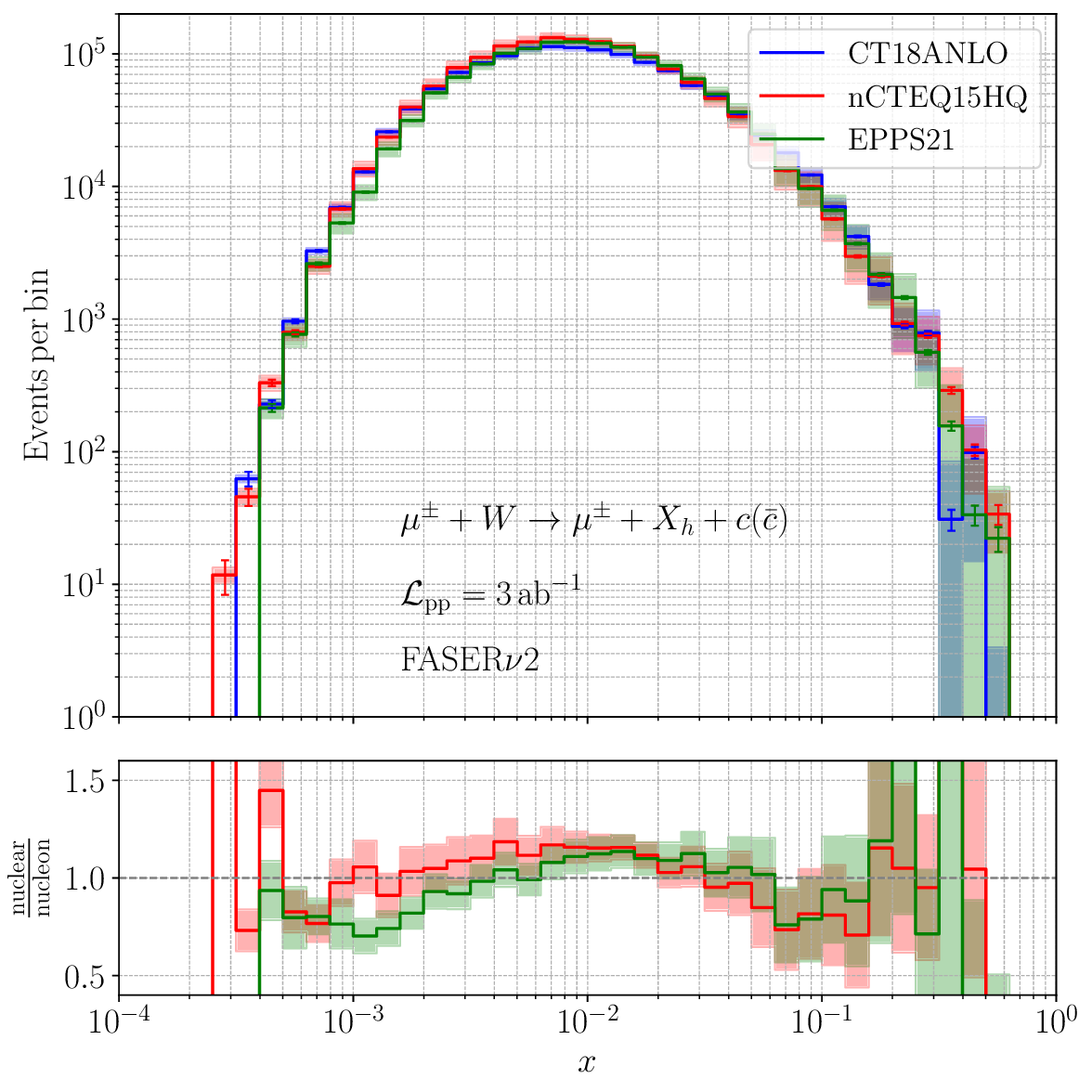}
	\includegraphics[width=0.5\textwidth]{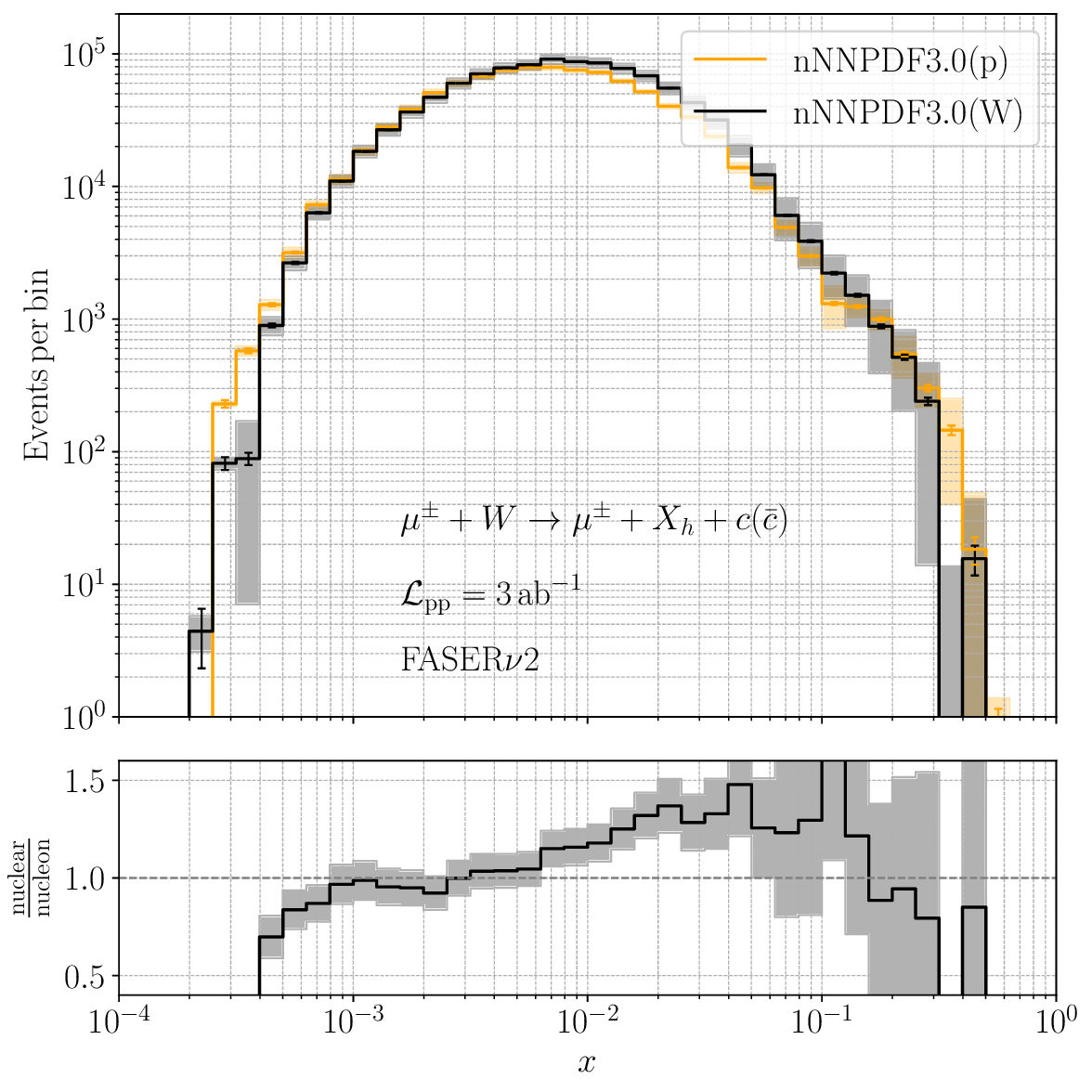}
    
			\end{tabular}
\caption{Predictions for the number of events for muon DIS at FASER$\nu2$ binned in $x$ in the inclusive case (upper panels)  and with  a tagged charm (lower panels) in the final state. Results derived assuming different parameterizations for the nPDFs. The predictions for the ratio between the nuclear and nucleon results are presented in the bottom panels of the plots.}
\label{fig:muonDIS_x_FASERnu2HL}
\end{figure}

\subsection{FASER$\nu 2$}
The results presented in the previous subsection strongly motivate to analyze how an upgraded experimental scenario can improve the constraining of nuclear effects in lepton - ion DIS at the LHC.  As discussed in Section \ref{sec:for},  FASER$\nu 2$ detector has been proposed to be installed in the Forward Physics Facility and operate during the HL-LHC era. The larger size of this future detector, and the huge increasing in the luminosity, are expected to increase the event rates and decrease the statistical uncertainties. Such expectation is confirmed by the results that we will show in what follows.

\begin{figure}[t]
	\centering
	\begin{tabular}{ccc}
	\includegraphics[width=0.5\textwidth]{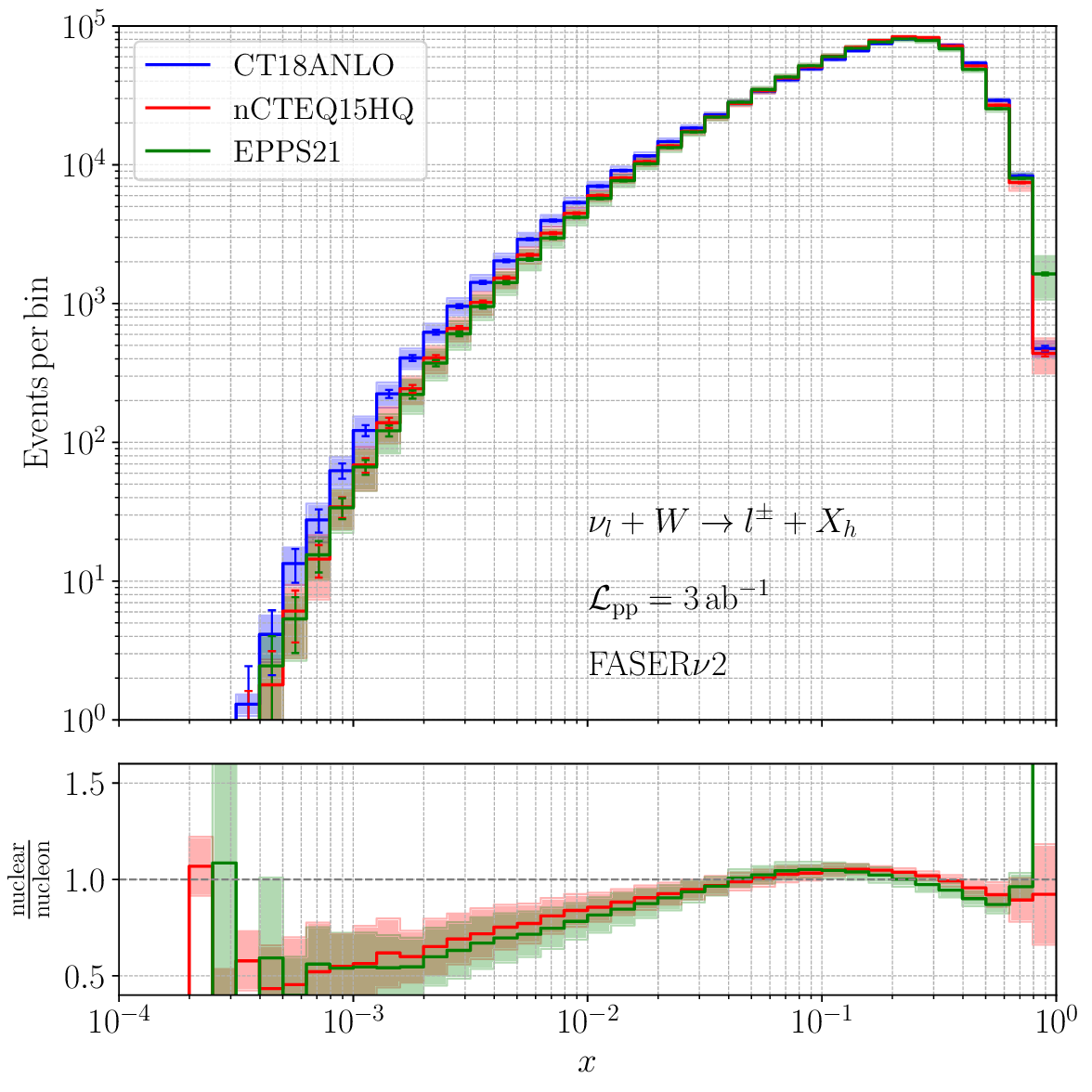}
	\includegraphics[width=0.5\textwidth]{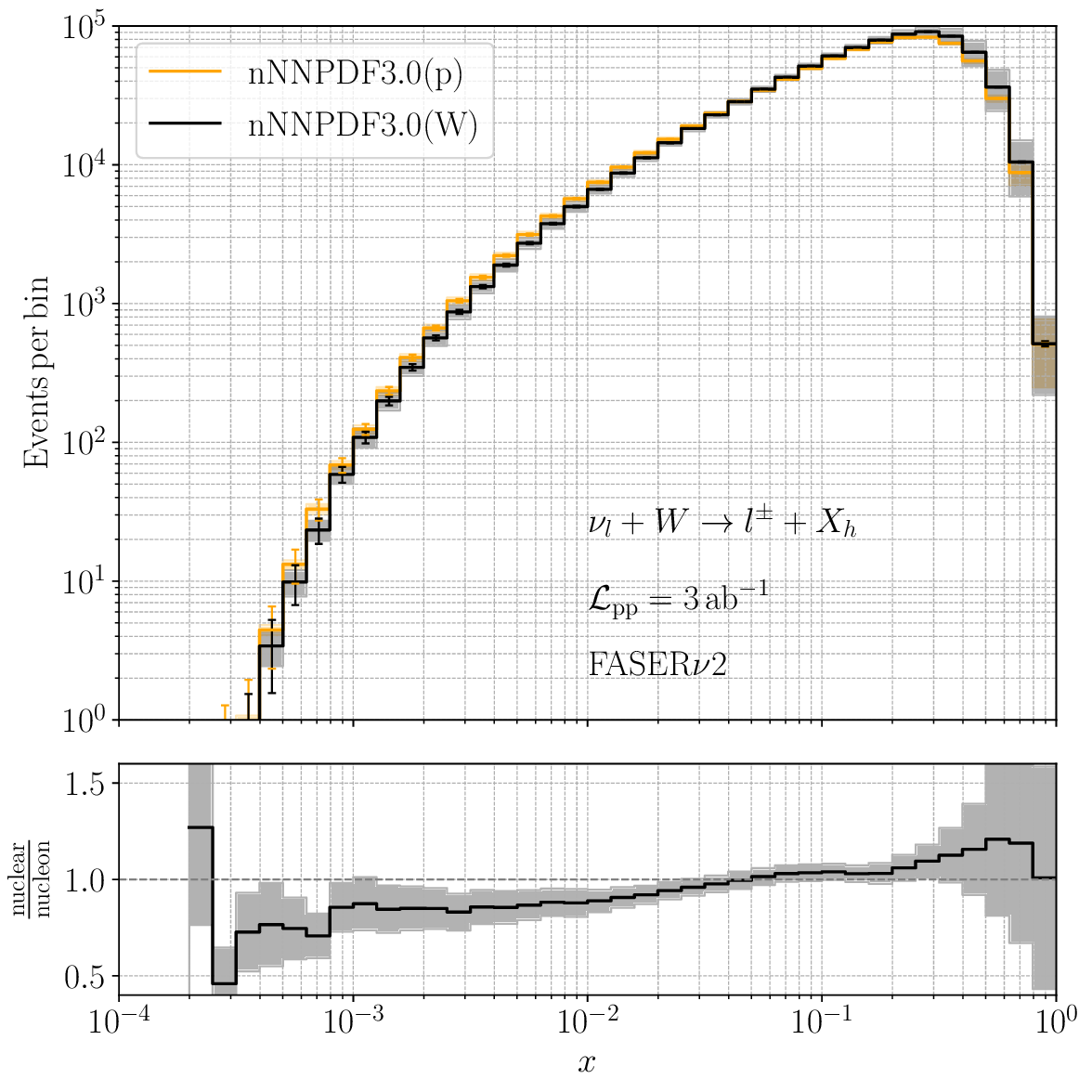} \\
    \includegraphics[width=0.5\textwidth]{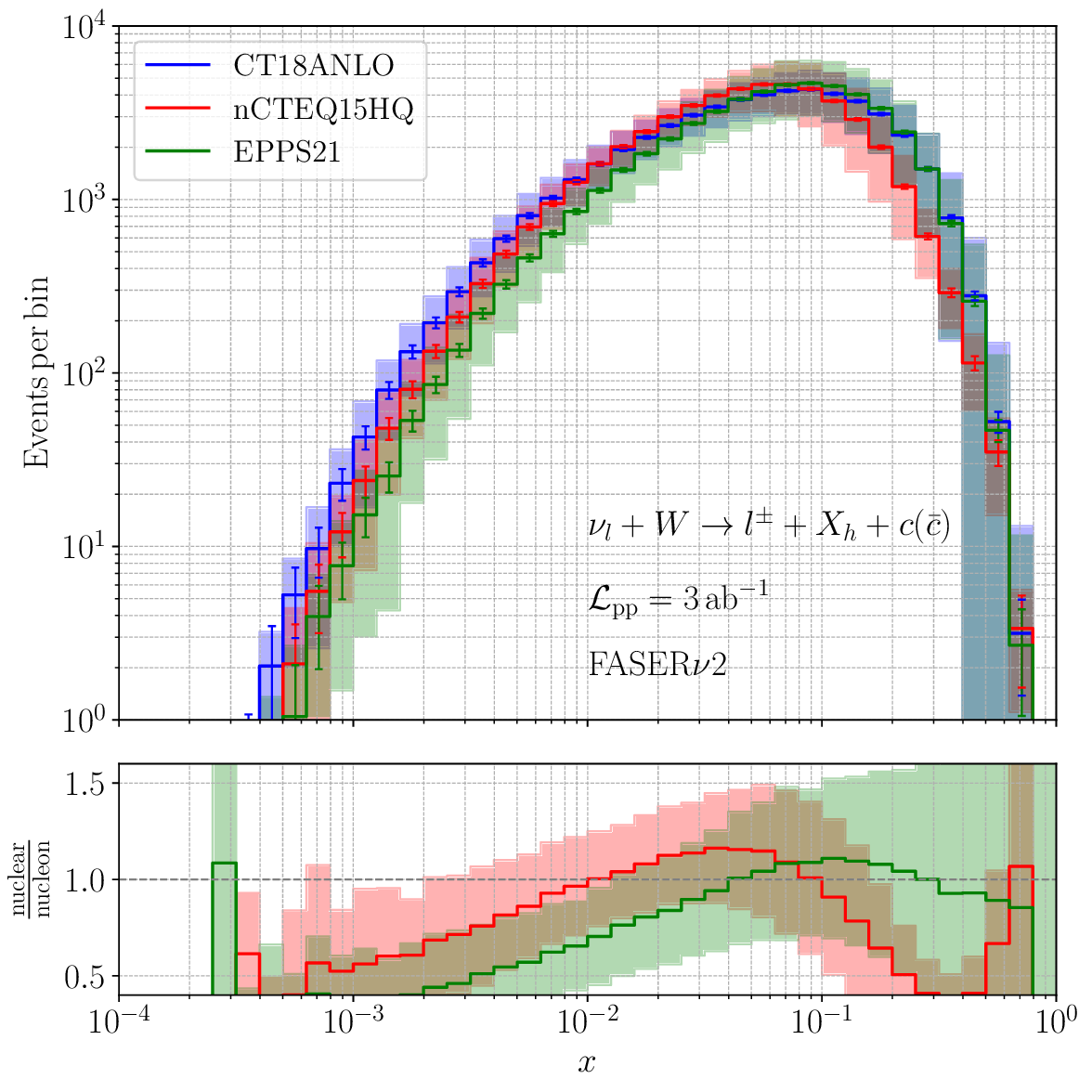}
	\includegraphics[width=0.5\textwidth]{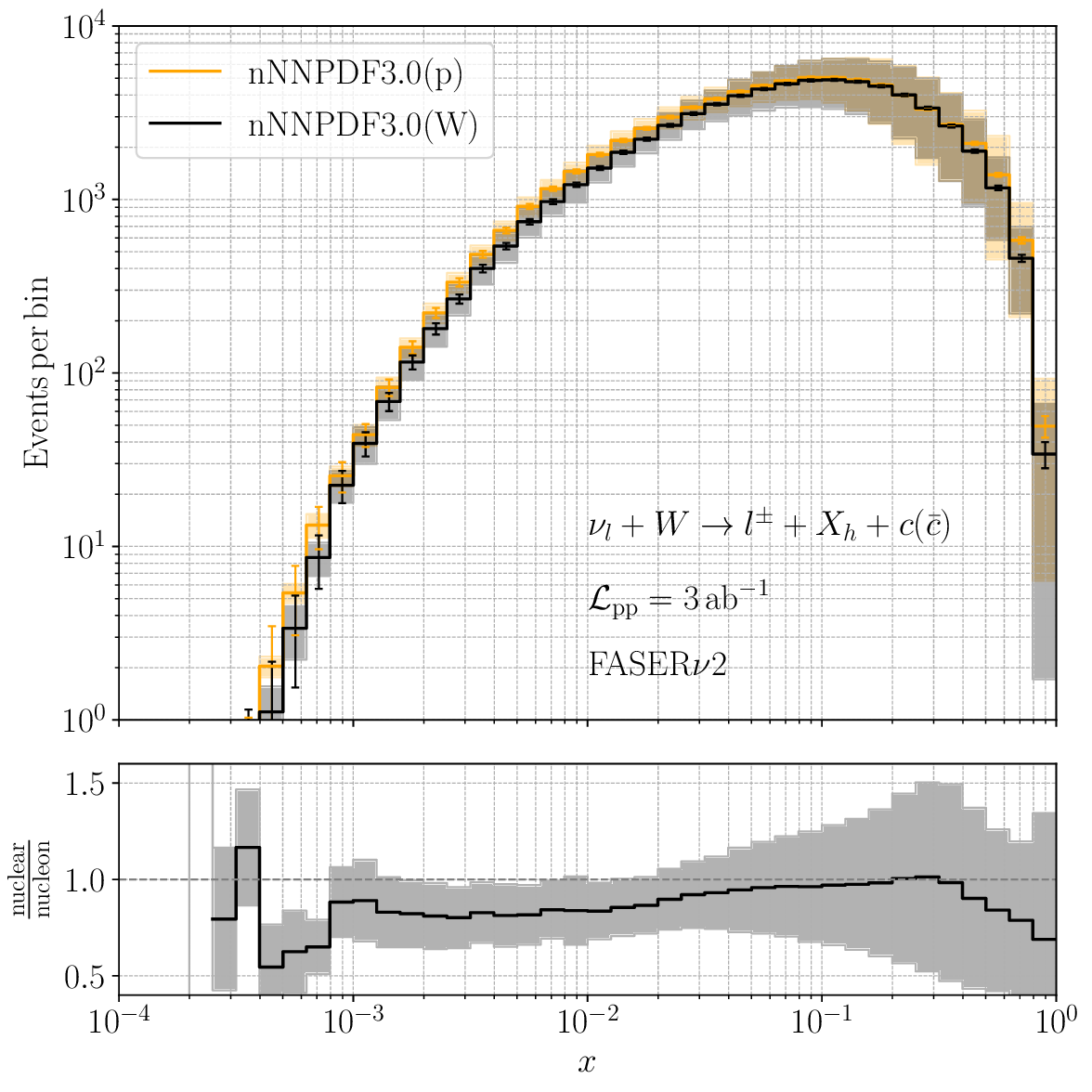}
    
			\end{tabular}
\caption{Predictions for the number of events for neutrino DIS at FASER$\nu2$ binned in $x$ in the inclusive case (upper panels)  and with  a tagged charm (lower panels) in the final state. Results derived assuming different parameterizations for the nPDFs. The predictions for the ratio between the nuclear and nucleon results are presented in the bottom panels of the plots. }
\label{fig:neutrinoDIS_x_FASERnu2HL}
\end{figure}

In Fig. \ref{fig:muonDIS_x_FASERnu2HL} we present our results for the  muon DIS events binned in $x$ at FASER$\nu 2$ considering an integrated luminosity of 3 ab$^{-1}$. We have an increasing in the event rates by approximately two orders of magnitude in comparison with the results for FASER$\nu$ in run 3. Moreover, the relative expected statistical error bars decrease by a factor of $\approx 10$, making the statistical uncertainties smaller than the current PDF uncertainty in the entire $x$ range considered. The differences between the predictions for the impact of the nuclear effects on inclusive and tagged charm events, derived using the distinct nPDFs, are also expected at FASER$\nu 2$. The same conclusion is also valid for neutrino DIS events, as demonstrated by the results presented in Fig. \ref{fig:neutrinoDIS_x_FASERnu2HL}. The results with distinct PDF parameterizations give similar predictions, but the expected statistical errors are smaller than the current PDF uncertainty, indicating that FASER$\nu 2$ can be a powerful tool to decrease the current nuclear PDFs uncertainties.

\begin{figure}[t]
	\centering
	\begin{tabular}{ccc}
	\includegraphics[width=0.5\textwidth]{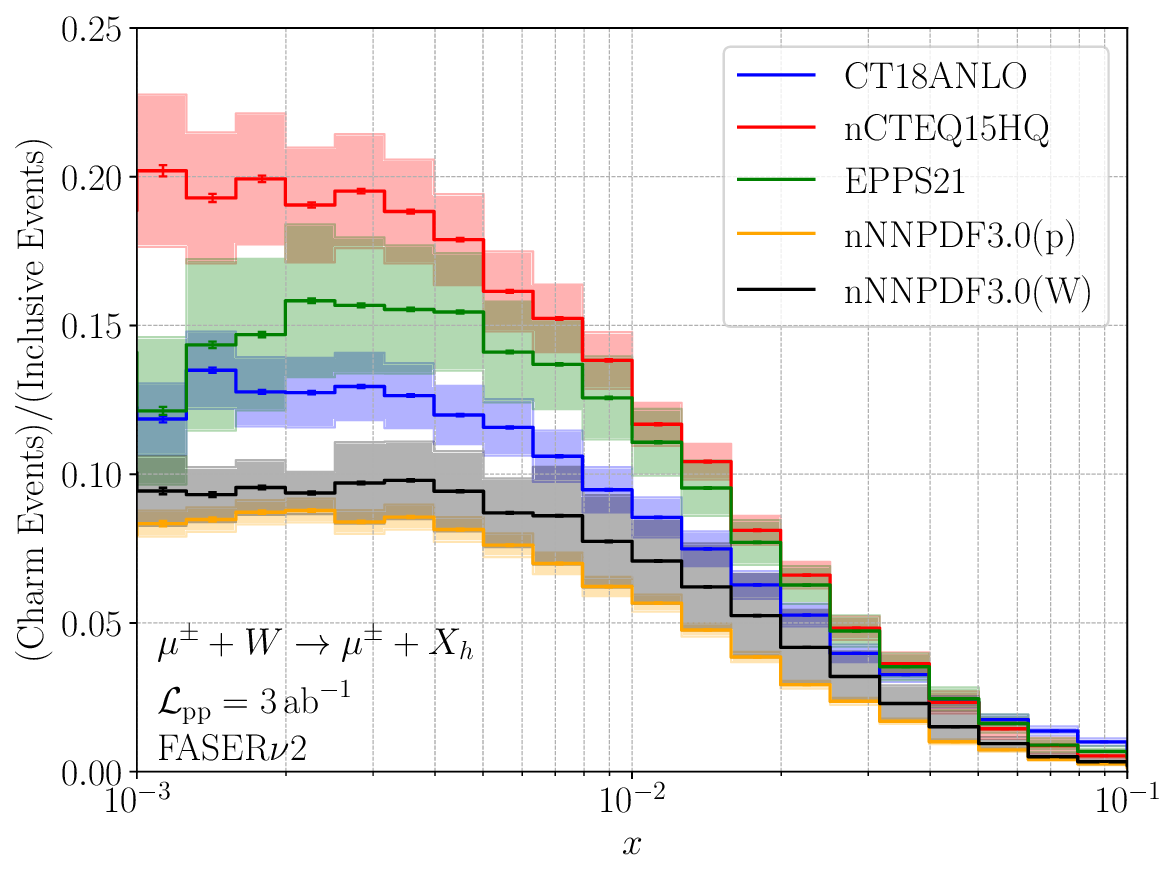}
    \includegraphics[width=0.5\textwidth]{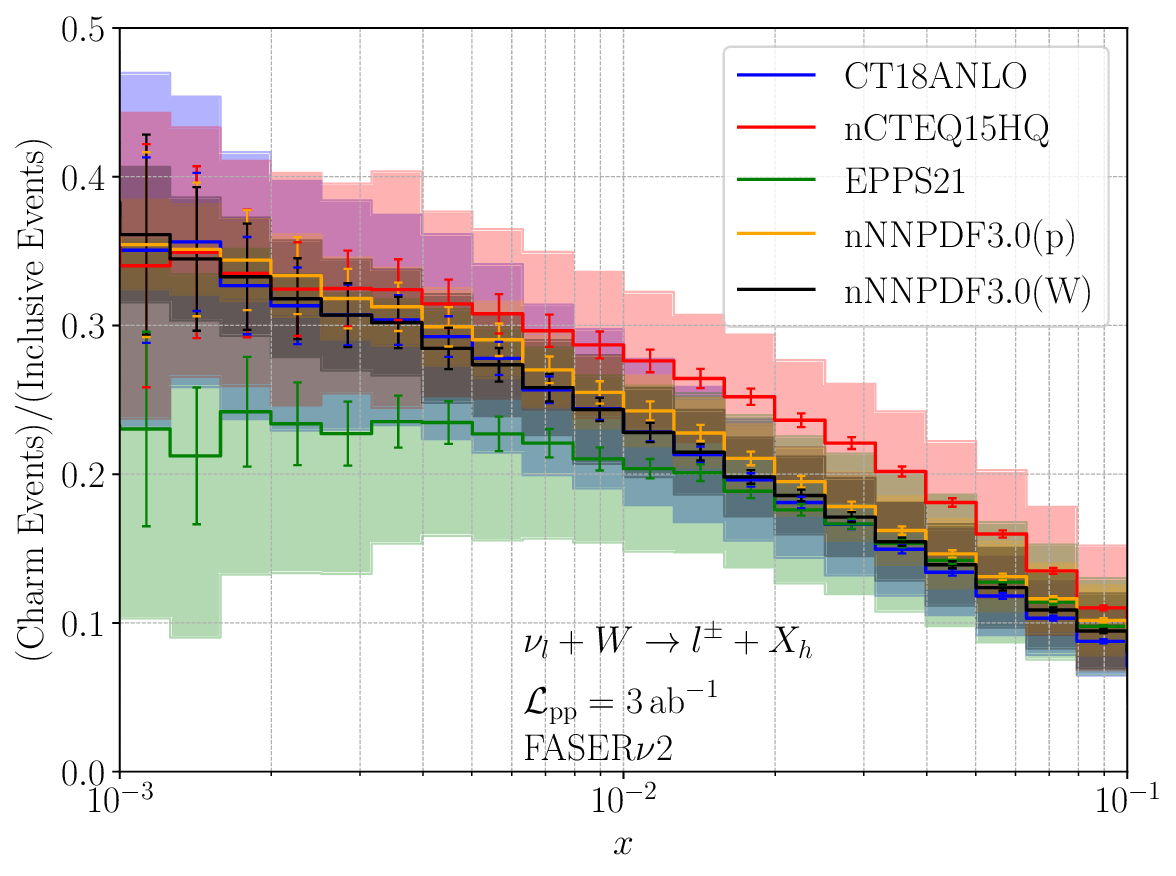}

			\end{tabular}
\caption{ Ratio between events with tagged charm hadron and inclusive for muon (left) and neutrino (right) DIS at FASER$\nu 2$ in the LHC high luminosity era. Results derived assuming different parameterizations for the nPDFs.}
\label{fig:Ratio_charm_total_FASERnu2_HL}
\end{figure}

Finally, in Fig. \ref{fig:Ratio_charm_total_FASERnu2_HL} we present our predictions for the ratio between charm tagged and inclusive events for muon (left panel) and neutrino (right panel) induced interactions. In comparison with the results shown in Fig. \ref{fig:Ratio_charm_total_FASERnu}, we have now that  the difference between the EPPS21 prediction for neutrino DIS and those derived using the other PDFs, will be larger than the expected statistical uncertainties. As a consequence, a future experimental analysis of the proposed ratio in $\mu W$ and $\nu W$ events, measured by the same detector, could be used to constrain the description of the nuclear effects, as well as to test the universality of the nPDFs.

\section{Summary}
\label{sec:sum}
The description of the nuclear effects in the parton distributions has been significantly improved during the last decades. However, despite the theoretical and experimental advances, the current difference between results for the nPDFs, derived by distinct groups that perform independent global analysis, is still non - negligible, and there are  tensions between different datasets, with the bulk of neutrino DIS data being incompatible with neutral current DIS data. These studies clearly indicate that new data is needed in order to improve our understanding of the nuclear effects (See, e.g., Ref. \cite{Khalek:2021ulf}). Such is one of the motivations for the construction of future electron - ion colliders \cite{eic,lhec,EicC}. However, the starting of the far - detector physics program at the LHC has demonstrated that the study of neutrino - ion and muon - ion interactions is also feasible in this collider, opening a new window into neutrino and hadronic physics. The possibility of measuring $\mu W$ and $\nu W$ DIS events in a same experiment will allow performing a precise test of the universality of the nPDFs. Moreover, it will allow us to improve the modeling of the nuclear cross - sections, which is fundamental for BSM searches of new particles at the far - forward detectors.  Motivated by these perspectives, in this paper we have investigated the impact of the nuclear effects on the cross - sections for neutrino - tungsten and muon - tungsten interactions at the FASER$\nu$ and FASER$\nu 2$ detectors. We have calculated the rates for inclusive and charm tagged events considering different parameterizations for the nuclear PDFs, which differ on its predictions for the amount of nuclear effects, especially at small - $x$ and in the strange and gluon distributions. Our results show that the predictions are  sensitive to the nPDF used as input in the calculations. We have proposed the analysis of the ratio between charm tagged and inclusive events, and demonstrated that a future experimental analysis of this quantity will be very useful to discriminate between the distinct descriptions already during the Run 3 of LHC using the FASER$\nu$ detector. Moreover, we have presented the corresponding predictions for the HL-LHC phase, assuming the operation of the  FASER$\nu 2$ detector, and demonstrated that the large amount of events will allow us to perform a very precise analysis of the lepton - DIS events at the LHC. Motivated by these results and those presented in Ref.~\cite{MammenAbraham:2024gun}, we intend to extend our analysis for the Future Circular Collider (FCC)~\cite{FCC:2018vvp} in a forthcoming study.

\begin{acknowledgments}
 The authors would like to thank the referee for the criticism that allow us to improve the manuscript and Juan Rojo for helpful comments and discussions. R. F. acknowledges support from the Conselho Nacional de Desenvolvimento Cient\'{\i}fico e Tecnol\'ogico (CNPq, Brazil), Grant No. 161770/2022-3. V.P.G. was partially supported by CNPq, FAPERGS and INCT-FNA (Process No. 464898/2014-5). D.R.G. was partially supported by CNPq and MCTI.

\end{acknowledgments}

\hspace{1.0cm}


\begin{thebibliography}{99}
\bibitem{Klasen:2023uqj}
M.~Klasen and H.~Paukkunen,
Ann. Rev. Nucl. Part. Sci. \textbf{74}, 49-87 (2024)



\bibitem{Eskola:2021nhw}
K.~J.~Eskola, P.~Paakkinen, H.~Paukkunen and C.~A.~Salgado,
Eur. Phys. J. C \textbf{82}, no.5, 413 (2022)


\bibitem{AbdulKhalek:2022fyi}
R.~Abdul Khalek, R.~Gauld, T.~Giani, E.~R.~Nocera, T.~R.~Rabemananjara and J.~Rojo,
Eur. Phys. J. C \textbf{82}, no.6, 507 (2022)



\bibitem{Duwentaster:2022kpv}
P.~Duwent{\"a}ster, T.~Je{\v{z}}o, M.~Klasen, K.~Kova{\v{r}}{\'\i}k, A.~Kusina, K.~F.~Muzakka, F.~I.~Olness, R.~Ruiz, I.~Schienbein and J.~Y.~Yu,
Phys. Rev. D \textbf{105}, no.11, 114043 (2022)


\bibitem{Muzakka:2022wey}
K.~F.~Muzakka, P.~Duwent{\"a}ster, T.~J.~Hobbs, T.~Je{\v{z}}o, M.~Klasen, K.~Kova{\v{r}}{\'\i}k, A.~Kusina, J.~G.~Morf{\'\i}n, F.~I.~Olness and R.~Ruiz, \textit{et al.}
Phys. Rev. D \textbf{106}, no.7, 074004 (2022)

\bibitem{Schienbein:2007fs}
I.~Schienbein, J.~Y.~Yu, C.~Keppel, J.~G.~Morfin, F.~Olness and J.~F.~Owens,
Phys. Rev. D \textbf{77}, 054013 (2008)




\bibitem{Paukkunen:2010hb}
H.~Paukkunen and C.~A.~Salgado,
JHEP \textbf{07}, 032 (2010)

\bibitem{Kovarik:2010uv}
K.~Kovarik, I.~Schienbein, F.~I.~Olness, J.~Y.~Yu, C.~Keppel, J.~G.~Morfin, J.~F.~Owens and T.~Stavreva,
Phys. Rev. Lett. \textbf{106}, 122301 (2011)

\bibitem{Paukkunen:2013grz}
H.~Paukkunen and C.~A.~Salgado,
Phys. Rev. Lett. \textbf{110}, no.21, 212301 (2013)



\bibitem{SNDLHC:2022ihg}
G.~Acampora \textit{et al.} [SND@LHC],
JINST \textbf{19}, no.05, P05067 (2024)

\bibitem{SNDLHC:2023pun}
R.~Albanese \textit{et al.} [SND@LHC],
Phys. Rev. Lett. \textbf{131}, no.3, 031802 (2023)

\bibitem{FASER:2023zcr}
H.~Abreu \textit{et al.} [FASER],
Phys. Rev. Lett. \textbf{131}, no.3, 3 (2023)

\bibitem{FASER:2024hoe}
R.~Mammen Abraham \textit{et al.} [FASER],
Phys. Rev. Lett. \textbf{133}, no.2, 021802 (2024)

\bibitem{FASER:2024ref}
R.~Mammen Abraham \textit{et al.} [FASER],
Phys. Rev. Lett. \textbf{134}, no.21, 211801 (2025)

\bibitem{Kling:2023tgr}
F.~Kling, T.~M{\"a}kel{\"a} and S.~Trojanowski,
Phys. Rev. D \textbf{108}, no.9, 095020 (2023)

\bibitem{John:2025qlm}
J.~John, F.~Kling, J.~Koorn, P.~Krack and J.~Rojo,
[arXiv:2507.06022 [hep-ph]].

\bibitem{Cruz-Martinez:2023sdv}
J.~M.~Cruz-Martinez, M.~Fieg, T.~Giani, P.~Krack, T.~M{\"a}kel{\"a}, T.~R.~Rabemananjara and J.~Rojo,
Eur. Phys. J. C \textbf{84}, no.4, 369 (2024)


\bibitem{Francener:2025pnr}
R.~Francener, V.~P.~Goncalves, F.~Kling, P.~Krack and J.~Rojo,
[arXiv:2506.13889 [hep-ph]].


\bibitem{Ariga:2023fjg}
A.~Ariga, R.~Balkin, I.~Galon, E.~Kajomovitz and Y.~Soreq,
Phys. Rev. D \textbf{109}, no.3, 035003 (2024)

\bibitem{MammenAbraham:2025gai}
R.~Mammen Abraham and M.~Fieg,
Phys. Rev. D \textbf{112}, no.1, 1 (2025)



\bibitem{Anchordoqui:2021ghd}
L.~A.~Anchordoqui, A.~Ariga, T.~Ariga, W.~Bai, K.~Balazs, B.~Batell, J.~Boyd, J.~Bramante, M.~Campanelli and A.~Carmona, \textit{et al.}
Phys. Rept. \textbf{968}, 1-50 (2022)

\bibitem{Feng:2022inv}
J.~L.~Feng, F.~Kling, M.~H.~Reno, J.~Rojo, D.~Soldin, L.~A.~Anchordoqui, J.~Boyd, A.~Ismail, L.~Harland-Lang and K.~J.~Kelly, \textit{et al.}
J. Phys. G \textbf{50}, no.3, 030501 (2023)



\bibitem{vanBeekveld:2024ziz}
M.~van Beekveld, S.~Ferrario Ravasio, E.~Groenendijk, P.~Krack, J.~Rojo and V.~S.~S{\'a}nchez,
Eur. Phys. J. C \textbf{84}, no.11, 1175 (2024)

\bibitem{Sabate-Gilarte:2023aeg}
M.~Sabate-Gilarte and F.~Cerutti,
JACoW \textbf{IPAC2023}, MOPL018 (2023)

\bibitem{Battistoni:2015epi}
G.~Battistoni, T.~Boehlen, F.~Cerutti, P.~W.~Chin, L.~S.~Esposito, A.~Fass{\`o}, A.~Ferrari, A.~Lechner, A.~Empl and A.~Mairani, \textit{et al.}
Annals Nucl. Energy \textbf{82}, 10-18 (2015)

\bibitem{Buckley:2014ana}
A.~Buckley, J.~Ferrando, S.~Lloyd, K.~Nordstr{\"o}m, B.~Page, M.~R{\"u}fenacht, M.~Sch{\"o}nherr and G.~Watt,
Eur. Phys. J. C \textbf{75}, 132 (2015)

\bibitem{Kling:2021gos}
F.~Kling and L.~J.~Nevay,
Phys. Rev. D \textbf{104}, no.11, 113008 (2021)

\bibitem{Buonocore:2023kna}
L.~Buonocore, F.~Kling, L.~Rottoli and J.~Sominka,
Eur. Phys. J. C \textbf{84}, no.4, 363 (2024)

\bibitem{Banfi:2023mhz}
A.~Banfi, S.~Ferrario Ravasio, B.~J{\"a}ger, A.~Karlberg, F.~Reichenbach and G.~Zanderighi,
JHEP \textbf{02}, 023 (2024)


\bibitem{FerrarioRavasio:2024kem}
S.~Ferrario Ravasio, R.~Gauld, B.~J{\"a}ger, A.~Karlberg and G.~Zanderighi,
[arXiv:2407.03894 [hep-ph]].

\bibitem{Bierlich:2022pfr}
C.~Bierlich, S.~Chakraborty, N.~Desai, L.~Gellersen, I.~Helenius, P.~Ilten, L.~L{\"o}nnblad, S.~Mrenna, S.~Prestel and C.~T.~Preuss, \textit{et al.}
SciPost Phys. Codeb. \textbf{2022}, 8 (2022)

\bibitem{Skands:2014pea}
P.~Skands, S.~Carrazza and J.~Rojo,
Eur. Phys. J. C \textbf{74}, no.8, 3024 (2014)

\bibitem{Hou:2019qau}
T.~J.~Hou, K.~Xie, J.~Gao, S.~Dulat, M.~Guzzi, T.~J.~Hobbs, J.~Huston, P.~Nadolsky, J.~Pumplin and C.~Schmidt, \textit{et al.}
[arXiv:1908.11394 [hep-ph]].



\bibitem{Khalek:2021ulf}
R.~A.~Khalek, J.~J.~Ethier, E.~R.~Nocera and J.~Rojo,
Phys. Rev. D \textbf{103} (2021) no.9, 096005



\bibitem{eic}
  D.~Boer, M.~Diehl, R.~Milner, R.~Venugopalan, W.~Vogelsang, D.~Kaplan, H.~Montgomery and S.~Vigdor {\it et al.},
  arXiv:1108.1713 [nucl-th];
  A.~Accardi, J.~L.~Albacete, M.~Anselmino, N.~Armesto, E.~C.~Aschenauer, A.~Bacchetta, D.~Boer and W.~Brooks {\it et al.},
Eur.\ Phys.\ J.\ A {\bf 52}, no. 9, 268 (2016);  E.~C.~Aschenauer {\it et al.},
  Rept.\ Prog.\ Phys.\  {\bf 82}, no. 2, 024301 (2019); R.~Abdul Khalek, A.~Accardi, J.~Adam, D.~Adamiak, W.~Akers, M.~Albaladejo, A.~Al-bataineh, M.~G.~Alexeev, F.~Ameli and P.~Antonioli, \textit{et al.}
Nucl. Phys. A \textbf{1026}, 122447 (2022); V.~D.~Burkert, L.~Elouadrhiri, A.~Afanasev, J.~Arrington, M.~Contalbrigo, W.~Cosyn, A.~Deshpande, D.~I.~Glazier, X.~Ji and S.~Liuti, \textit{et al.}
Prog. Part. Nucl. Phys. \textbf{131}, 104032 (2023); 
R.~Abir, I.~Akushevich, T.~Altinoluk, D.~P.~Anderle, F.~P.~Aslan, A.~Bacchetta, B.~Balantekin, J.~Barata, M.~Battaglieri and C.~A.~Bertulani, \textit{et al.}
[arXiv:2305.14572 [hep-ph]].

\bibitem{lhec}
  J.~L.~Abelleira Fernandez {\it et al.}  [LHeC Study Group Collaboration],
  J.\ Phys.\ G {\bf 39}, 075001 (2012); P.~Agostini {\it et al.},
J. Phys. G \textbf{48}, no.11, 110501 (2021).



\bibitem{EicC}
D.~P.~Anderle, V.~Bertone, X.~Cao, L.~Chang, N.~Chang, G.~Chen, X.~Chen, Z.~Chen, Z.~Cui and L.~Dai, \textit{et al.}
Front. Phys. (Beijing) \textbf{16}, no.6, 64701 (2021).



\bibitem{MammenAbraham:2024gun}
R.~Mammen Abraham, J.~Adhikary, J.~L.~Feng, M.~Fieg, F.~Kling, J.~Li, J.~Pei, T.~R.~Rabemananjara, J.~Rojo and S.~Trojanowski,
JHEP \textbf{01} (2025), 094


\bibitem{FCC:2018vvp}
A.~Abada \textit{et al.} [FCC],
Eur. Phys. J. ST \textbf{228} (2019) no.4, 755-1107



 
\end{thebibliography}
\end{document}